\RequirePackage[2018-12-01]{latexrelease}
\documentclass{rQUF2e}

\usepackage{epstopdf}% To incorporate .eps illustrations using PDFLaTeX, etc.
\usepackage{subfigure}% Support for small, `sub' figures and tables

\theoremstyle{plain}

\theoremstyle{definition}

\theoremstyle{remark}

\usepackage{amsmath, amsfonts, amsbsy, amssymb}
\usepackage{graphicx}
\usepackage{multirow}
\usepackage{booktabs}
\usepackage{adjustbox,subcaption,url}

\begin{document}

%\jvol{00} \jnum{00} \jyear{2014} \jmonth{October}

\title{A mixture transition distribution approach to portfolio optimization }

\author{R. DE BLASIS$\dagger$ and L. GALATI$^{\ast}$$\ddagger$\thanks{$^\ast$Corresponding author.
Email: luca.galati@unibo.it} and F. PETRONI${\S}$\\
\affil{$\dagger$Marche Polytechnic University, Piazzale Martelli 8, Ancona, AN, 60121, Italy\\
$\ddagger$University of Bologna, Via Capo di Lucca 34, Bologna, BO, 40126, Italy\\
$\S$University of Chieti-Pescara “G. D'Annunzio”, Viale Pindaro 42, Pescara, PE, 65127, Italy} 
\received{\today}
}

\maketitle

\begin{abstract}

Understanding the dependencies among financial assets is critical for portfolio optimization. Traditional approaches based on correlation networks often fail to capture the nonlinear and directional relationships that exist in financial markets. In this study, we construct directed and weighted financial networks using the Mixture Transition Distribution (MTD) model, offering a richer representation of asset interdependencies. We apply local assortativity measures—metrics that evaluate how assets connect based on similarities or differences—to guide portfolio selection and allocation.
Using data from the Dow Jones 30, Euro Stoxx 50, and FTSE 100 indices constituents, we show that portfolios optimized with network-based assortativity measures consistently outperform the classical mean-variance framework. Notably, modalities in which assets with differing characteristics connect enhance diversification and improve Sharpe ratios. The directed nature of MTD-based networks effectively captures complex relationships, yielding portfolios with superior risk-adjusted returns.
Our findings highlight the utility of network-based methodologies in financial decision-making, demonstrating their ability to refine portfolio optimization strategies. This work thus underscores the potential of leveraging advanced financial networks to achieve enhanced performance, offering valuable insights for practitioners and setting a foundation for future research.

\end{abstract}

\begin{keywords}
Financial Networks; Mixture Transition Distribution; Multivariate Markov chain; Portfolio allocation; Network mixing; Network assortativity.
\end{keywords}

\begin{classcode}C45; C58; C61; G11; G17.\end{classcode}

\section{Introduction}\label{intro}

Understanding the intricate interconnections among financial assets and their dynamic behaviors is critical for optimizing portfolio management in financial markets. Network theory has recently emerged as a powerful tool for analyzing such relationships, using correlation matrices as the foundation for constructing financial networks with nodes and links represented by securities and their interconnections, respectively. These networks often undergo filtering processes, such as the minimum spanning tree \citep[e.g.,][]{mantegna1999HierarchicalStructureFinancial}, the planar maximally filtered graph \citep[e.g.,][]{tumminello2005ToolFilteringInformation}, the triangulated maximally filtered graph \citep[e.g.,][]{massara2017NetworkFilteringBig}, or correlation threshold \citep[e.g.,][]{ricca2024PortfolioOptimizationNetwork}, to manage their density. However, these methods have limitations in fully capturing the nonlinear dependencies among financial assets. Recent advancements, such as the Mixture Transition Distribution (MTD) model proposed by \cite{damico2023MixtureTransitionDistribution}, address these shortcomings by modeling nonlinear relationships and producing directed and weighted networks. Building on these developments, this paper introduces a novel approach to portfolio optimization by adopting an MTD-based financial network to represent the complex dependencies among financial assets.

One of the earliest applications of network theory to financial markets was introduced by \cite{mantegna1999HierarchicalStructureFinancial}, who analyzed the returns of the Dow Jones Industrial Average (DJIA) and Standard \& Poor’s 500 (S\&P 500) indices. By deriving a measure of distance from return correlations, \cite{mantegna1999HierarchicalStructureFinancial} constructed networks using the minimum spanning tree (MST) approach, which reduces the number of links to $n - 1$, where $n$ represents the number of assets. This method revealed that interconnected stocks are often clustered by industry. Building on this, \cite{onnela2003AssetTreesAsset} proposed dynamic asset graphs by creating networks of 477 stocks traded on the New York Stock Exchange (NYSE). While still relying on return correlations, they retained only the closest $n - 1$ nodes, resulting in a graph structure rather than a tree. Extending this line of research, \cite{tse2010NetworkPerspectiveStock}. developed a comprehensive network of U.S. stocks with 19,807 nodes, connecting stocks if their correlation exceeded a specified threshold. More recently, \cite{lyocsa2012StockMarketNetworks} applied dynamic conditional correlation (DCC) methods to build networks for the S\&P 500 constituents and \cite{guo2022MultiLikelihood} introduced a maximum likelihood estimation approach to determine stock-specific thresholds, finding that traditional moving window approaches offered greater robustness in capturing industry clusters.

\cite{tumminello2005ToolFilteringInformation}, instead, introduced a heuristic algorithm to construct the Planar Maximally Filtered Graph (PMFG) as an alternative method for filtering correlation matrices. This approach was applied to analyze the topological characteristics of PMFG networks constructed from the returns of the 300 largest stocks listed on the New York Stock Exchange (NYSE) during the period 2001–2003, examining different time horizons. More recently, \cite{massara2017NetworkFilteringBig} proposed the Triangulated Maximally Filtered Graph (TMFG), an efficient triangulation-based algorithm for filtering correlation matrices. This method has demonstrated versatility in various financial contexts. For instance, \cite{deblasis2024InformationFlow} examined the issue in \cite{galati2024MarketBehavior} using the TMFG approach to study the cryptocurrency network of the FTX exchange during the collapse of its native token, FTT. Their analysis leveraged vertex centrality measures to explore how network structures respond to significant financial shocks, providing insights into the adaptive dynamics of financial networks in periods of crisis.

Moving beyond correlation-based methods, alternative frameworks have emerged to capture more nuanced relationships. \cite{billio2012EconometricMeasuresConnectedness} used principal component analysis and pairwise Granger-causality to construct networks among hedge funds, mutual funds, and financial institutions, offering insights into interdependencies within the financial sector. \cite{yang2014CointegrationAnalysisInfluence} analyzed cointegration relationships among global stock indices, enabling the construction of directed networks to represent causal links. However, these methods often lack the ability to assign weights to edges. Addressing this limitation, \cite{su2022} combined Granger-causality and cointegration tests to create directed and weighted networks using a sliding window methodology. Similarly, \cite{diebold2014} constructed weighted and directed networks of U.S. financial institutions using vector autoregression (VAR) variance decomposition, while \cite{yang2023} built sovereign default networks, leveraging centrality measures to explore their role in driving currency risk premia. A further innovation by \cite{chen2021} involved constructing multi-layered networks that integrated correlation, grey relational analysis, and maximum information coefficients.

However, very recently \cite{damico2023MixtureTransitionDistribution} introduced a novel methodology for constructing stock networks by modeling stock returns as a multivariate Markov chain using the Mixture Transition Distribution (MTD) model. Initially proposed by \cite{raftery1985ModelHighOrderMarkov} to handle high-order Markov chains and later extended by \cite{ching2002MultivariateMarkovChain} to a multivariate context, the MTD framework has seen various financial applications, including stock valuation, price discovery, and credit risk \citep{damico2023MixtureTransitionDistribution}. In their study, \cite{damico2023MixtureTransitionDistribution} employed the multivariate MTD to derive a connectedness matrix, which serves as the adjacency matrix for network construction and captures dependencies that extend beyond simple linear correlations. They thus demonstrated the potential of this approach by applying it to the Dow Jones 30 constituents, highlighting its ability to model asymmetric dependencies among stocks. Their findings showcased how MTD-based networks allow for the calculation of both in-degree and out-degree centrality, underscoring the practical applicability of the methodology in real-world financial scenarios. Building on this foundation, we expand the use of MTD networks to portfolio optimization. We do this by incorporating local assortativity measures, providing a new lens through which to evaluate asset relationships and enhance portfolio optimization.

\cite{newman2003MixingPatternsNetworks} is among the first to study assortative mixing in networks, namely the propensity of nodes in networks to connect with other vertices that are like (or unlike) them in some way. In their study, the authors proposed several models and measures to probe that assortative mixing gives important insights about the networks' design and functionality, and indeed documented that it is a pervasive phenomenon in many real-world networks. This was taken as inspiration by \cite{piraveenan2008LocalAssortativenessScalefree,piraveenan2010LocalAssortativenessScalefree}, who introduced a measure of local assertiveness that quantifies the level of assortative mixing for individual nodes. Further recent studies introduced different local assortativity measures, such as the ones by \cite{peel2018MultiscaleMixingPatterns}, \cite{pigorsch2022AssortativeMixingWeighted}, and \cite{sabek2023LocalAssortativityWeighted}, and provided disparate real-world applications. 
In financial market networks, where securities are represented as vertices, having nodes less assortative (or nodes more disassortative) indicates that the financial assets within the market exhibit diverse characteristics—a favorable scenario for portfolio optimization. The usefulness of these measures in a portfolio management context was indeed understood by a very recent study by \cite{ricca2024PortfolioOptimizationNetwork}, who extended the local assortativity measure of \cite{piraveenan2008LocalAssortativenessScalefree} to weighted networks and applied it to portfolio selection in three large financial markets.

In a similar vein, we take advantage of these measures to demonstrate their effectiveness in guiding the selection and weighting of assets within portfolios. In this study, we extend the \cite{piraveenan2010LocalAssortativenessScalefree} measure, weighted as proposed by \cite{ricca2024PortfolioOptimizationNetwork}, to directed networks constructed using the MTD model mentioned above. We then compare it against two other widely used local assortativity measures \citep[i.e.,][]{peel2018MultiscaleMixingPatterns,sabek2023LocalAssortativityWeighted}, framing the analysis as a comparative “horse race" to highlight their relative strengths and applicability. We employ the Max Quadratic Utility and Sharpe Ratio optimization methods—established standards in modern portfolio theory \citep{markowitz1952modern}—to apply the model of \cite{damico2023MixtureTransitionDistribution} to real-world financial data from major equity indices constituents, including the Dow Jones 30, Euro Stoxx 50, and FTSE 100. Through an out-of-sample empirical analysis, we compare the effectiveness of our network-based assortativity measures against both the classical mean-variance framework and other correlation-based network approaches. Our findings demonstrate the superiority of MTD-network-based metrics in capturing the complex dependencies among assets, leading to portfolios with improved risk-adjusted returns.

This study contributes to the literature on network-based portfolio optimization by introducing a novel approach and several extensions. First, we move beyond the traditional use of correlation-based networks and employ the mixture transition distribution model, which captures nonlinear relationships and generates directed networks, allowing for a richer representation of dependencies among financial assets. This application of the MTD model to portfolio optimization marks an advancement in the field. Second, we extend the framework proposed by \cite{piraveenan2010LocalAssortativenessScalefree} by incorporating directed networks, thereby enriching the local assortativity analysis and offering new insights into the directional influence among assets. Third, while taking inspiration from \cite{ricca2024PortfolioOptimizationNetwork}, we distinguish our approach by adopting different modern portfolio theory optimization standards—specifically, the Max Quadratic Utility and Sharpe Ratio methodologies proposed by \cite{markowitz1952modern}. Unlike \cite{ricca2024PortfolioOptimizationNetwork}, who focus on return maximization subject to a pre-defined Value at Risk (VaR), our approach prioritizes the dual objectives of maximizing returns and minimizing risk within a robust theoretical framework. Finally, we provide empirical evidence demonstrating the practical utility of MTD-network-based assortativity measures in portfolio management. By consistently outperforming the gold-standard methods of modern portfolio theory, our results offer valuable insights for practitioners seeking novel and more efficient models for portfolio optimization, underscoring the potential of network theory as a powerful tool for achieving superior risk-adjusted returns.

The paper is organized as follows. Section \ref{model} outlines the methodology, including the MTD model, directed networks, and assortativity measures. Section \ref{application} describes the data and sample selection of the financial application and presents the empirical findings, comparing network-based measures to the Markowitz benchmark. Section \ref{conclusion} concludes with key insights and implications for portfolio management.

\section{Model}\label{model}

\subsection{Construction of the financial network}
As shown in \cite{damico2023MixtureTransitionDistribution}, it is possible to build a weighted and directed financial network based on the MTD model from \cite{raftery1985ModelHighOrderMarkov}, extended to a multivariate setting by \cite{ching2002MultivariateMarkovChain}. To this extent, let us consider a multivariate sequence of random variables $\bm{S}=(S_t^{(i)}, \forall i \in N=\{1,2,...,n\})$, defined on a probability space $(\Omega,\mathcal{F},\mathbb{P})$ and taking values on the same finite state space $\mathcal{Z}$. In addition, let us assume that the sequence respects the following multivariate Markov property:
\begin{align}\label{eq:MarkovProperty}
\mathbb{P}(S_{t+1}^{(j)}=k|(S_t^{(1)}=h_t^{(1)},S_{t-1}^{(1)}=h_{t-1}^{(1)},...,S_0^{(1)}=h_0^{(1)}),...,\nonumber\\
(S_t^{(n)}=k_t^{(n)},S_{t-1}^{(n)}=h_{t-1}^{(n)},..., S_0^{(n)}=h_0^{(n)}))\\
=\mathbb{P}(S_{t+1}^{(j)}=k|S_{t}^{(1)}=h_t^{(1)},...,S_{t}^{(n)}=h_t^{(n)}).\nonumber
\end{align}
Property (\ref{eq:MarkovProperty}) states that the probability of being in state $k$ at time $t+1$ for the $j$-th series depends only on the state $ h_t^{(1)},...,h_t^{(n)} $ occupied by all series at time $t$. Modeling property (\ref{eq:MarkovProperty}) becomes infeasible when the number of series increases because we would need to take into account all transitions from starting states to ending states combinations. In order to reduce the number of parameters, we can employ the MTD model which applies a convex linear combination to the transition probability matrices from one series to another. Specifically, we can compute the probability distribution of series $j$ at time $t+1$ as:
\begin{equation}\label{eq:MTDproperty}
    \mathbf{D}^{(j)}(t+1)=\sum_{i=1}^{n}\mathbf{D}^{(i)}(t)\cdot \lambda_{ij}\cdot \mathbf{P}^{(i,j)},
\end{equation}
where $ \mathbf{D}^{j}(t):=[D_{1}^{(j)},\ldots,D_{z}^{(j)}] $, $ D_{h}^{(j)}(t):=\mathbb{P}(S^{(j)}_t=h)$, and $\mathbf{P}^{(i,j)}=(p^{(i,j)}_{hk})_{h,k\in\mathcal{Z}}$ is the transition probability matrix containing the probabilities of moving from state $h$ in series $i$ to state $k$ in series $j$.

The scalar parameters of the linear combination, $\lambda_{ij}$, are subject to the following constraints:
\begin{equation}\label{eq:LambdaConditions}
    \sum_{i=1}^{n}\lambda_{ij}=1,\forall j \in N, \quad \lambda_{ij}\ge 0.
\end{equation}
They measure the degree of dependence among the different series of the systems. Specifically, when considering financial returns series, equation (\ref{eq:MTDproperty}) states that the probability for a price change in series $j$ of being in a specific state (e.g., negative, positive, or null) is a linear combination of all the transition probabilities from each series initial states to the arrival state in series $j$. In other words, large values of $\lambda_{ij}$ weights indicate a strong influence from returns of series $i$ to returns of series $j$. For the estimation procedure to compute the transition probabilities $\mathbf{P}^{\beta,\alpha}$ and the parameters $\lambda_{\beta,\alpha}$ we refer the reader to \cite{damico2023MixtureTransitionDistribution}.

Now, considering that $j\in N$, we obtain $n$ different equations (\ref{eq:MTDproperty}) for each arrival return series. Therefore, we can build a matrix of $\lambda_{ij}$ weights,
\begin{equation}\label{eq:lambdaMatrix}
\bf{\Lambda}=
\begin{pmatrix}
    \lambda_{1,1} & \lambda_{1,2} & \cdots & \lambda_{1,n}\\
	\lambda_{2,1} & \lambda_{2,2} & \cdots & \lambda_{2,n}\\
	\vdots        & \vdots        & \ddots & \vdots\\
	\lambda_{n,1} & \lambda_{n,2} & \cdots & \lambda_{n,n}\\
\end{pmatrix}.
\end{equation}
Then, from the matrix $\bm{\Lambda}$ we build a directed and weighted financial network described by a graph $G=(N,E)$ with $n=|N|$ nodes and $e=|E|$ edges. Therefore, the graph $G$ has adjacency matrix, $\bm{A}$, with elements
\begin{equation}\label{eq:AdjacencyMatrix}
    a_{ij}=
    \begin{cases}
        1 & \text{ if } \lambda_{ij}>0 \text{ and } i\ne j\\
        0 & \text{otherwise}
    \end{cases},
\end{equation}
and weighted adjacency matrix, $\bm{W}$, with elements
\begin{equation}\label{eq:WAdjacencyMatrix}
    w_{ij}=
    \begin{cases}
        \lambda_{ij} & \text{ if } i\ne j\\
        0 & \text{otherwise}
    \end{cases}.
\end{equation}
Both conditions in (\ref{eq:AdjacencyMatrix}) and (\ref{eq:WAdjacencyMatrix}) eliminate the self-loops in the graph.

For each node $i=1,...,n$, we compute some node's characteristics. We denote by $d_i^{in}=\sum_j a_{ji}$ and $d_i^{out}=\sum_j a_{ij}$ the in- and out-degrees of node $i$, respectively. We identify $N(i)$ as the set of neighbors of node $i$. In particular, we consider only the direct successors of node $i$; thus, the neighbors' cardinality is equal to $d_i^{out}$. Additionally, we indicate by $s_i^{in}=\sum_j w_{ji}$ and $s_i^{out}=\sum_j w_{ij}$ the in- and out-strength of node $i$, respectively. 

\subsection{Network assortativity}
The network assortativity measures the correlation between the distribution of some characteristics of the graph on pairs of adjacent nodes \citep{newman2003MixingPatternsNetworks}. A common approach in the literature is to employ the excess degree or excess strength as node characteristics and compute a simple or weighted correlation between them. For our purpose, we consider the excess in- and out-strength as $es_i^{in}=s_i^{in}-w_ji$ and $es_i^{out}=s_i^{out}-w_ij$, respectively. Further, we employ the weighted correlation approach. Thus, following \cite{pigorsch2022AssortativeMixingWeighted}, the general formula of the global assortativity for directed and weighted graphs is:
\begin{equation}\label{eq:global}
    \rho_g(m_1,m_2)=\frac{\sum_{i,j}w_{ij}es_i^{m_1}es_j^{m_2}-\Omega^{-1}\left(\sum_{i,j}w_{ij}es_i^{m_1}\right)\left(\sum_{i,j}w_{ij}es_j^{m_2}\right)}{\sqrt{\sum_{i,j}w_{ij}(es_i^{m_1})^2-\Omega^{-1}\left(\sum_{i,j}w_{ij}es_i^{m_1}\right)^2}\sqrt{\sum_{i,j}w_{ij}(es_j^{m_2})^2-\Omega^{-1}\left(\sum_{i,j}w_{ij}es_j^{m_2}\right)^2}},
\end{equation}
where the couple $(m_1,m_2)$, with $m_1,m_2\in\{in, out\}$, defines the mode of the assortativity, $es_i^{m_1}$ is the excess in- or out-strength of the source node $i$ when considering edge $(i,j)$ (similarly for the target node $j$), and $\Omega=\sum_{i,j}w_{ij}$.

The global assortativity is a single coefficient describing the entire network. However, to properly diversify a financial portfolio, we need a single coefficient for each node. To this extent, we analyze three different measures of local assortativity.

As a first measure, we propose an extension of the local assortativity introduced by \cite{piraveenan2008LocalAssortativenessScalefree,piraveenan2010LocalAssortativenessScalefree} to directed and weighted networks. This local assortativity measures the contribution that each node makes to the global assortativity. Given a source node $i$, we compute the assortativity as the weighted correlation between the source node $i$ and its neighbors $N(i)$. In our directed network, we assume that the neighbors are all direct successors of node $i$, i.e., all nodes reached with an out-edge from node $i$. Therefore, the local assortativity can be computed as
\begin{equation}\label{eq:Extended_Piraveenan}
   \rho_i(m_1,m_2)=\frac{\sum_{j\in N(i)}w_{ij}es_i^{m_1}es_j^{m_2}-\Omega^{-1}\left(\sum_{j\in N(i)}w_{ij}es_i^{m_1}\right)\left(\sum_{i,j}w_{ij}es_j^{m_2}\right)}{\sqrt{\sum_{i,j}w_{ij}(es_i^{m_1})^2-\Omega^{-1}\left(\sum_{i,j}w_{ij}es_i^{m_1}\right)^2}\sqrt{\sum_{i,j}w_{ij}(es_j^{m_2})^2-\Omega^{-1}\left(\sum_{i,j}w_{ij}es_j^{m_2}\right)^2}},
\end{equation}
where the first and second terms in the numerator measure the contribution of node $i$ to the cross-product of excess strengths and the average source strength, respectively. The denominator is the scaling factor which is unchanged to ensure that the sum of the local assortativities is equal to the global measure. 
 
A second measure has been introduced by \cite{sabek2023LocalAssortativityWeighted}. The authors start with the definition of the local assortativity on the edges of the network,
\begin{equation}\label{eq:sabekedge}
    \rho_{e_{ij}}(m_1,m_2)=\frac{w_{ij}(es_i^{m_1}-\mu_i^{m_1})(es_j^{m_2}-\mu_j^{m_2})}{\Omega\sigma_i^{m_1}\sigma_j^{m_2}},
\end{equation}
where $\mu_i^{m_1}=\sum_{ij}w_{ij}es_i^{m_1}$ is the weighted mean excess (in- or out-) strength of the source nodes on all edges (similarly for $\mu_j^{m_2}$), and $\sigma_i^{m_1}$ and $\sigma_j^{m_2}$ are the weighted standard deviations of the excess (in- or out-) strengths over the whole network.

Then, the assortativity of a node can be computed as the sum of the edge assortativity of all the node's neighbors,
\begin{equation}\label{eq:sabeknode}
    \rho_i(m_1,m_2)=\sum_{j\in N(i)}\rho_{e_ij}(m_1,m_2)
\end{equation}

Finally, a third measure has been proposed by \cite{peel2018MultiscaleMixingPatterns}. The authors argue that considering only the neighbors of a node to compute the local assortativity encounters problems. In particular, for nodes with low degree, we would compute the assortativity on a small sample, thus providing a poor estimate of the node's mixing preference. They propose a solution by reweighting the edges in the network based on how local they are to the node of interest $l$. Therefore, formula (\ref{eq:sabeknode}) can be rewritten as
\begin{equation}\label{eq:peel}
    \rho_l(m_1,m_2)=\sum_{ij}w_{\alpha}(i;l)\frac{w_{ij}(es_i^{m_1}-\bar{es_i^{m_1}})(es_j^{m_2}-\bar{es_j^{m_2}})}{s_i^{out}\sigma_i^{m_1}\sigma_j^{m_2}},
\end{equation}
where $w_{\alpha}(i;l)$ is a distribution over the nodes. \cite{peel2018MultiscaleMixingPatterns} suggest to employ the personalized PageRank vector as a specific distribution, i.e., the stationary distribution of a random walk with restart to node $l$ with probability $(1-\alpha)$. The random walker can move on the network $G$ jumping from node $i$ to node $j$ with probability $w_{ij}/s_i^{out}$. As for the choice of $\alpha$, the authors propose to integrate $w_{\alpha}(i;l)$ over all possible values, thus obtaining a multi-scale distribution $w_{multi}(i;l)=\int_0^1 w_{\alpha}(i;l)d\alpha$ to substitute for $w_{\alpha}(i;l)$ in (\ref{eq:peel}). Using this distribution, we compute a multiscale local measure that captures the assortativity of a node across all scales. Contrary to the previous two local measures, we observe that the sum of these local assortativities is not equal to the global assortativity.

\subsection{Portfolio optimization}
Following \cite{ricca2024PortfolioOptimizationNetwork}, we extend two classical portfolio problems, i.e., the max quadratic utility and max Sharpe ratio problems, with the addition of the information derived from the excess strength assortativity. 

Let $\bm{\mu}$ and $\bm{\Sigma}$ be the assets' returns vector and covariance matrix with dimensions $n$ and $n\times n$, respectively, with $n$ equal to the number of assets. Let $\bm{x}$ be the vector of portfolio weights. The extended max quadratic utility optimization can be written
\begin{equation}
    \max \quad \bm{\mu}\bm{x}-\frac{\delta}{2}\bm{x}'\bm{\Sigma}\bm{x}-R,
\end{equation}
while the extended max Sharpe optimization takes the following form
\begin{equation}
    \max \quad \frac{\bm{\mu}\bm{x}}{\bm{x}'\bm{\Sigma}\bm{x}}-R,
\end{equation}
where $R$ is the total portfolio assortativity and it is subtracted from the objective function because we aim at selecting stocks that are disassortative to help diversify the asset allocation.

The portfolio assortativity can be computed as a simple or weighted sum of the selected stocks' assortativities:
\begin{align}
    R=\bm{\rho}\bm{x} \label{eq:weightedAss} \\ 
    R=\bm{\rho}\bm{y}, \label{eq:simpleAss}
\end{align}
where $\bm{\rho}$ is the vector of assortativities, $y_i=1$ if $x_i>0$ and $y_i=0$ if $x_i=0$.

The constraints of the optimization are:
\begin{align}
    & \sum_{i=1,...,n} x_i=1 \\ \label{eq:const_1}
    & x_i\ge 0, \quad i=1,...,n \\ \label{eq:const_2}
    & \gamma y_i \le x_i \le y_i, \quad i=1,...,n, \\ \label{eq:const_3}
    & y_i \in \{0,1\} \\ 
\end{align}
where equation (\ref{eq:const_1}) represents the budget constraints, equation (\ref{eq:const_2}) only allows for long positions in the portfolio, and equation (\ref{eq:const_3}) establishes a logical dependency between variables $x$ and $y$ with lower bound $\gamma$.

\section{Financial application}\label{application}

\subsection{Data and sample}\label{data}

This study uses daily closing price data sourced from Datastream, a product of Refinitiv, an LSEG business. The dataset encompasses stocks included in three major indices: the Dow Jones 30 (DJ30), Euro Stoxx 50 (STOXX50), and FTSE 100 (FTSE100), providing a cross-country comparison across the United States, Europe, and the United Kingdom, respectively.\footnote{Consistent with \cite{ricca2024PortfolioOptimizationNetwork}, the choice of these indices allows for a broad representation of market dynamics across different regions, providing a robust foundation for the portfolio optimization framework explored in this research.} For all indices, weekends and public holidays have been excluded from the analysis to ensure consistency in trading days. 

Consistent with \cite{ricca2024PortfolioOptimizationNetwork}, we assess the performance of our portfolios using a widely recognized rolling time window method, which permits portfolio re-balancing at regular intervals throughout the holding period. Specifically, we use a 90-day in-sample window and a 30-day out-of-sample period, with re-balancing occurring every 30 trading days. The 20-year sample spans from December 2004 to December 2024, comprising more than 5,000 (160) observations (rolling windows) for the constituents of each equity index and including 28 stocks for the DJ30, 46 stocks for the STOXX50, and 78 stocks for the FTSE100.
% The details of the datasets for the three indices are as follows:
% \begin{enumerate}
%     \item DJ30: The sample spans from December $17^{\text{th}}$, 2004, to December $16^{\text{th}}$, 2024, comprising 5034 observations. The analysis includes 28 stocks with a total of 164 rolling windows.
%     \item STOXX50: The sample covers December $14^{\text{th}}$, 2004, to December $13^{\text{th}}$, 2024, with 5126 observations. The dataset consists of 46 stocks and 167 rolling windows.
%     \item FTSE100: The sample extends from December $17^{\text{th}}$, 2004, to December $13^{\text{th}}$, 2024, comprising 5053 observations. It includes 78 stocks and 165 rolling windows.
% \end{enumerate}

We compute the logarithmic returns for each stock in the dataset to capture relative price changes, as follows:
\begin{equation}
    r_{i,t} = \ln\left(\frac{P_{i,t}}{P_{i,t-1}}\right),
\end{equation}
where $P_{i,t}$ represents the closing price of stock $i$ on day $t$, and $P_{i,t-1}$ denotes the closing price of the same stock on the previous trading day. We then apply the MTD model to portfolio optimization and compute the average expected returns for each portfolio over the rolling windows, as well as the portfolio returns' standard deviation and Sharpe Ratio using a constant risk-free rate of return equal to zero for convenience, consistent with \cite{ricca2024PortfolioOptimizationNetwork}.

Table \ref{tab:ass_stats} summarizes the network assortativity of our sample, broken down into three financial markets—Dow Jones 30, Euro Stoxx 50, and FTSE100—and three different assortativity measures (Extended Piraveenan et al., 2010; Sabek and Pigorsch, 2023; Peel et al., 2018). We consider four assortativity modalities: in-in, in-out, out-in, and out-out. Each modality provides insights into the connectivity preferences of nodes within the network. The measures include global assortativity ($\rho_g$), the proportion of assortative nodes ($P(\rho_i>0)$), the average assortativity of assortative nodes ($\overline{(\rho_i)+}$), and the average assortativity of disassortative nodes ($\overline{(\rho_i)-}$).

Global assortativity ($\rho_g$) determines whether the network exhibits assortativity or disassortativity overall. For the in-in modality, $\rho_g$ is consistently positive across all markets and measures, with values ranging between 0.015 and 0.022. This indicates a weak assortative tendency, where nodes that are highly influenced by others show a slight preference for connecting with similarly influenced nodes. In contrast, the in-out, out-in, and out-out modalities consistently exhibit negative $\rho_g$, indicating disassortative tendencies. For instance, $\rho_g$ in the in-out modality ranges from -0.059 to -0.093, reflecting that nodes influenced by others tend not to connect with nodes that strongly influence others. Similarly, the out-in modality has $\rho_g$ values ranging from -0.086 to -0.127, while the out-out modality shows values from -0.045 to -0.069, highlighting that highly influential nodes tend to avoid connecting with similarly influential nodes.

The proportion of assortative nodes ($P(\rho_i>0)$),  the probability of having a $\rho_g$ greater than zero, captures the percentage of assortative nodes that are in the network analyzed. This varies across modalities and measures. In the in-in modality, this proportion is highest, ranging from 0.552 (Sabek and Pigorsch, 2023, Dow Jones 30) to 0.644 (Extended Piraveenan et al., 2010, FTSE100), suggesting that a significant number of nodes individually exhibit assortative behavior, even when the global assortativity is weak. Conversely, in the in-out modality, the proportion of assortative nodes is moderate, with values between 0.342 and 0.427, reflecting a lower tendency for individual nodes to exhibit assortative connections. The out-in modality displays even lower proportions, with values ranging from 0.143 (Peel et al., 2018, Eurostoxx 50) to 0.654 (Extended Piraveenan et al., 2010, FTSE100), underscoring the variability introduced by the different measurement approaches. Finally, in the out-out modality, the proportion of assortative nodes is relatively balanced, with values between 0.363 and 0.526, indicating that assortative connections are neither dominant nor negligible.

The average assortativity of assortative nodes ($\overline{(\rho_i)_+}$) further reveals the magnitude of assortativity for positively assortative nodes. In the in-in modality, Peel et al. (2018) assortativity measure consistently reports high values, ranging from 0.172 to 0.239, while the Sabek and Pigorsch (2023) measure shows nearly negligible values, with averages between 0.003 and 0.012.\footnote{Albeit a comparison between local assortativity measures is by definition not viable.} Similar patterns are observed in the in-out modality, where the Peel et al. (2018) metric exhibits values from 0.146 to 0.196, again demonstrating stronger assortative tendencies compared to the near-zero averages reported by Sabek and Pigorsch. For the out-in modality, the Peel et al. (2018) measure reports averages ranging from 0.217 to 0.253, which are markedly higher than those observed in the other measures. Lastly, in the out-out modality, Peel et al. (2018) assortativity measure again reports the highest averages, ranging from 0.202 to 0.244, highlighting the tendency of this metric to capture stronger assortative nodes across modalities.

In contrast, the average assortativity of disassortative nodes ($\overline{(\rho_i)_-}$) quantifies the magnitude of disassortativity among negatively assortative nodes. The Peel et al. (2018) measure consistently reports the strongest disassortativity across all modalities. For instance, in the in-in modality, its values range from -0.174 to -0.208, compared to the Extended Piraveenan et al. (2010) measure, which values range from -0.078 to -0.143, and Sabek and Pigorsch (2023), with averages close to zero (-0.004 to -0.013). Similar trends are observed in the in-out modality, where Peel et al.'s (2018) assortativity reports values between -0.162 and -0.233, and in the out-in modality, where their values are significantly more disassortative, ranging from -0.391 to -0.459. The out-out modality follows the same pattern, with the Peel et al. (2018) reporting values between -0.244 and -0.293, indicating that disassortative nodes in this modality exhibit the strongest tendency to connect with dissimilar nodes.

Comparisons across markets reveal distinct patterns as well. The FTSE100 consistently exhibits higher proportions of assortative nodes across all modalities, suggesting a more cohesive network structure. The Dow Jones 30 shows moderate assortativity, generally weaker than the FTSE100 but stronger than the Eurostoxx 50. The Eurostoxx 50, on the other hand, consistently demonstrates the weakest assortativity and the strongest disassortativity, reflecting a more fragmented or heterogeneous market structure. These results illustrate significant variations in network connectivity dynamics across financial markets, as well as the impact of different assortativity measures on the interpretation of node behavior.

\begin{table}
    \centering
    \caption{Local assortativity analysis of the three markets for different assortativity modes. The averages of the rolling networks are reported. $\rho_g$ is the average global assortativity. $P(\rho_i>0)$ is the proportion of assortative nodes. $\overline{(\rho_i)_+}$ and $\overline{(\rho_i)_-}$ are the local assortativity averages of the assortative and disassortative nodes, respectively.}
    \label{tab:ass_stats}
    \begin{adjustbox}{max width=\textwidth}
    \begin{tabular}{lrrrlrrrlrrr}
    \toprule
     & \multicolumn{3}{c}{Extended Piraveenan et al. (2010)} && \multicolumn{3}{c}{Sabek and Pigorsch (2023)} && \multicolumn{3}{c}{Peel et al. (2018)} \\\cmidrule{2-4}\cmidrule{6-8}\cmidrule{10-12}
     & DJ30 & STOXX50& FTSE100 && DJ30 & STOXX50 & FTSE100 && DJ30 & STOXX50 & FTSE100 \\
    \midrule
    \multicolumn{12}{l}{\textit{Panel A: (in, in)}} \\
    $\rho_g$ & 0.021 & 0.022 & 0.015 && 0.021 & 0.022 & 0.015 && 0.021 & 0.022 & 0.015  \\
    $P(\rho_i>0)$ & 0.628 & 0.642 & 0.644 & & 0.552 & 0.563 & 0.575 & & 0.578 & 0.599 & 0.608 \\
    $\overline{(\rho_i)_+}$  & 0.083 & 0.060 & 0.043 & & 0.012 & 0.006 & 0.003 & & 0.239 & 0.221 & 0.172 \\
    $\overline{(\rho_i)_-}$  & -0.143 & -0.108 & -0.078 & & -0.013 & -0.007 & -0.004 & & -0.208 & -0.194 & -0.174 \\
    \midrule
    \multicolumn{12}{l}{\textit{Panel B: (in, out)}} \\
    $\rho_g$  & -0.088 & -0.093 & -0.059 & & -0.088 & -0.093 & -0.059 && -0.088 & -0.093 & -0.059\\
    $P(\rho_i>0)$  & 0.427 & 0.401 & 0.416 & & 0.409 & 0.379 & 0.406 & & 0.379 & 0.342 & 0.382 \\
    $\overline{(\rho_i)_+}$  & 0.104 & 0.080 & 0.057 & & 0.011 & 0.006 & 0.003 & & 0.196 & 0.175 & 0.146 \\
    $\overline{(\rho_i)_-}$  & -0.083 & -0.057 & -0.042 & & -0.013 & -0.007 & -0.003 & & -0.233 & -0.201 & -0.162 \\
    \midrule
    \multicolumn{12}{l}{\textit{Panel C: (out, in)}} \\
    $\rho_g$  & -0.086 & -0.127 & -0.093 & & -0.086 & -0.127 & -0.093 & & -0.086 & -0.127 & -0.093 \\
    $P(\rho_i>0)$  & 0.616 & 0.616 & 0.654 & & 0.390 & 0.353 & 0.387 & & 0.198 & 0.143 & 0.170 \\
    $\overline{(\rho_i)_+}$ & 0.014 & 0.006 & 0.004 & & 0.014 & 0.008 & 0.004 && 0.244 & 0.253 & 0.217 \\
    $\overline{(\rho_i)_-}$  & -0.031 & -0.018 & -0.011 & & -0.014 & -0.009 & -0.005 & & -0.459 & -0.446 & -0.391 \\
    \midrule
    \multicolumn{12}{l}{\textit{Panel D: (out, out)}}\\
    $\rho_g$  & -0.069 & -0.051 & -0.045 & & -0.069 & -0.051 & -0.045 & & -0.069 & -0.051 & -0.045 \\
    $P(\rho_i>0)$  & 0.395 & 0.363 & 0.390 & & 0.490 & 0.524 & 0.526 & & 0.415 & 0.446 & 0.468 \\
    $\overline{(\rho_i)_+}$  & 0.018 & 0.010 & 0.006 & & 0.010 & 0.005 & 0.003 & & 0.244 & 0.213 & 0.202 \\
    $\overline{(\rho_i)_-}$  & -0.016 & -0.008 & -0.005 & & -0.014 & -0.008 & -0.005 & & -0.293 & -0.271 & -0.244 \\
    \bottomrule
    \end{tabular}
    \end{adjustbox}
\end{table}

\begin{figure}
    \centering
    \includegraphics[width=\textwidth]{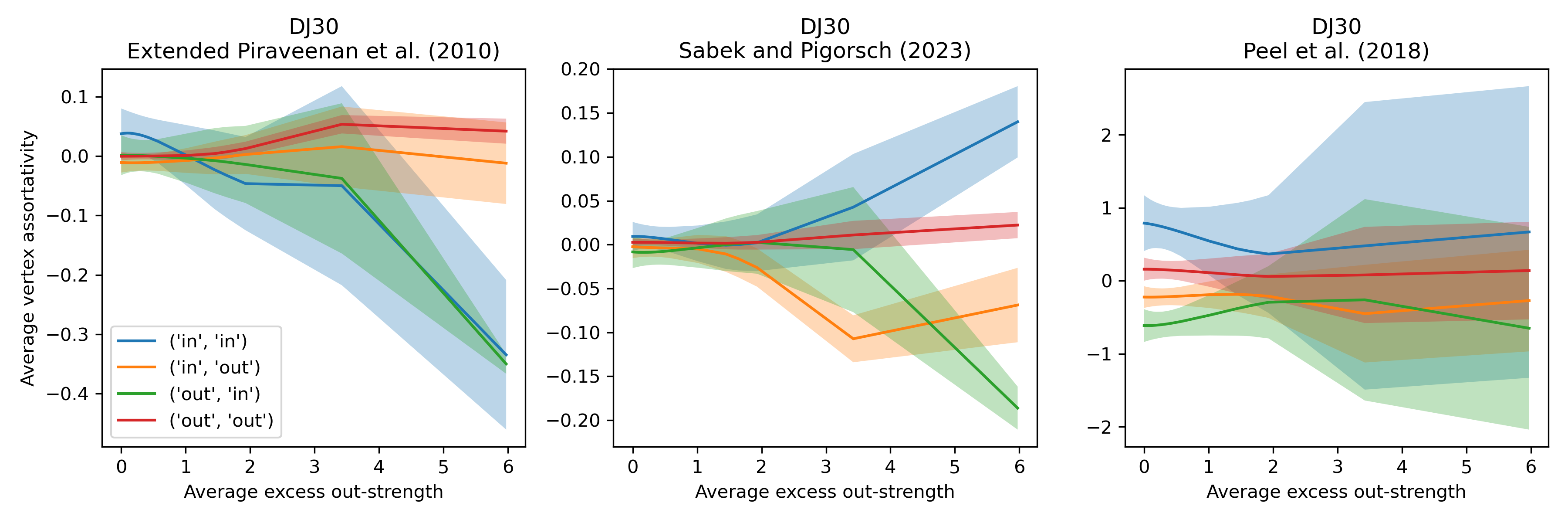}
    \includegraphics[width=\textwidth]{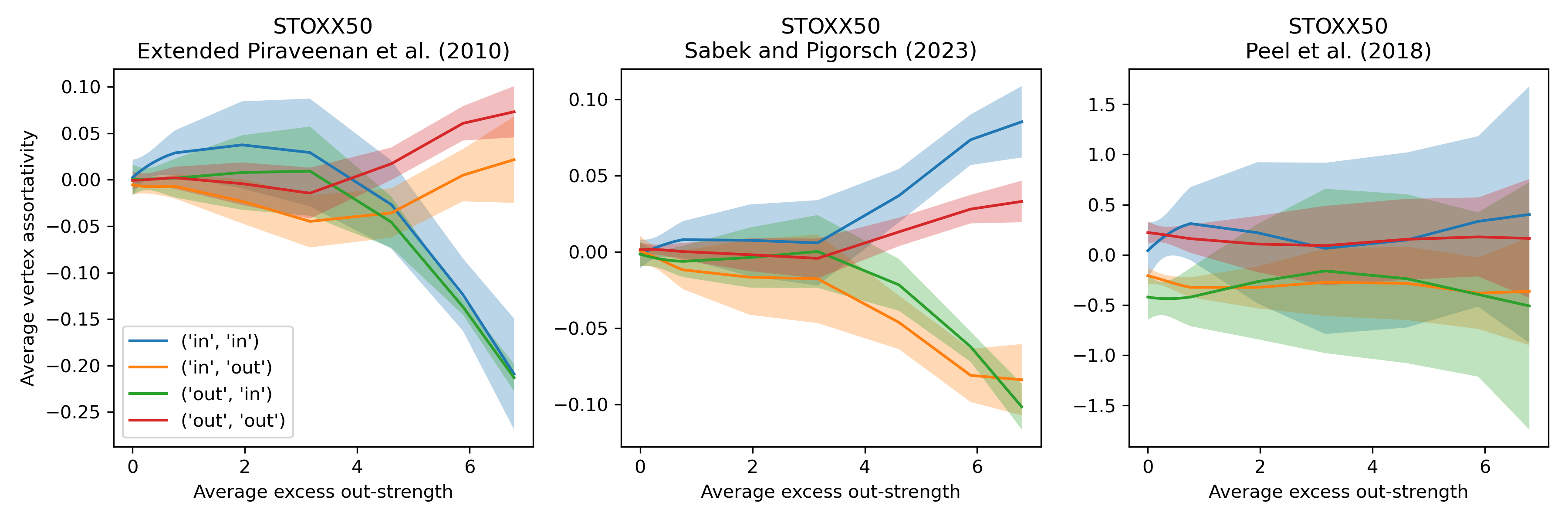}
    \includegraphics[width=\textwidth]{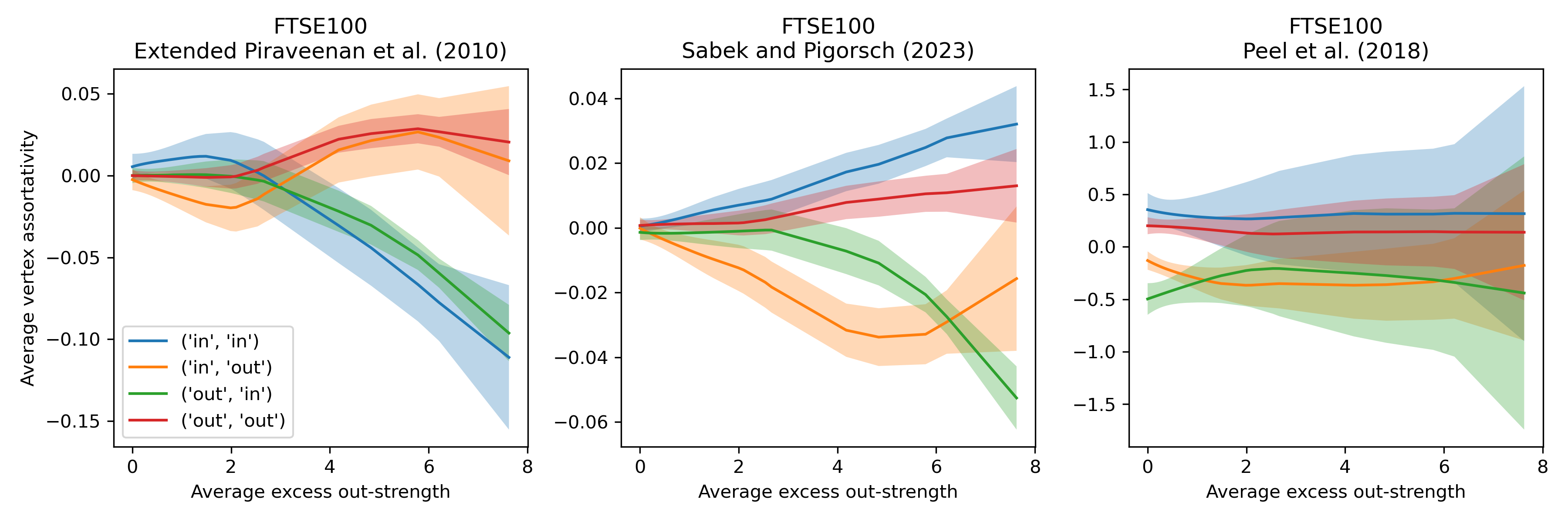}
    \caption{Relationship between vertex assortativity and excess out-strenght. The profiles are generated by applying loess regression to smooth the data, with the shaded area representing the 95 per cent confidence intervals.}
    \label{fig1}
\end{figure}

Figure \ref{fig1} illustrates the relationship between average vertex assortativity and the average excess out-strength of nodes for the three financial markets analyzed, using the three local assortativity measures. The results are stratified by four assortativity modalities: in-in, in-out, out-in, and out-out. Each plot provides a dynamic perspective on how assortative behavior evolves as the average excess out-strength of the nodes increases, with the x-axis measured in units corresponding to node weight.

The Extended Piraveenan et al. (2010) measure reveals distinct patterns across all three markets. For the in-in modality, average vertex assortativity starts positive, indicating a preference for nodes with high average excess out-strength to connect with other nodes similarly influenced. However, as the excess out-strength of the nodes increases, this positive tendency diminishes significantly, reflecting a reduction in assortative clustering among nodes with higher weights. For the in-out and out-in modalities, assortativity remains consistently negative across all levels of excess out-strength, indicating persistent disassortativity, where nodes with higher excess out-strength (e.g., influential nodes) avoid connecting with nodes of opposing roles, such as highly influenced nodes. The out-out modality exhibits near-zero assortativity across all markets, suggesting weak or negligible clustering tendencies between nodes of similar high excess out-strength. %These trends are consistent across the markets, although the Eurostoxx 50 exhibits slightly weaker assortativity in the in-in modality compared to the FTSE100 and Dow Jones 30.

The Sabek and Pigorsch (2023) measure exhibits subtler trends. For the in-in modality, assortativity remains weakly positive across all markets, with a slight increase observed as the average excess out-strength of the nodes rises. This suggests a mild preference for connections among highly influenced nodes. For the in-out and out-in modalities, average assortativity remains slightly negative throughout, with disassortative tendencies becoming marginally stronger as the excess out-strength increases. The out-out modality transitions from near-zero to slightly positive assortativity at higher levels of excess out-strength, reflecting a weak preference for connections between influential nodes. Among the markets, the FTSE100 consistently exhibits stronger in-in assortativity than the Dow Jones 30 and Eurostoxx 50, while the trends for the other modalities remain relatively similar across the three markets.

The Peel et al. (2018) measure shows flat dynamic patterns across all modalities and markets, signaling a more stable type of assortativity measure. For the in-in modality, assortativity starts strongly positive and increases significantly as the average excess out-strength rises, indicating a clear clustering tendency among highly influenced nodes. The in-out and out-in modalities show pronounced negative trends, with disassortativity intensifying as the excess out-strength increases, highlighting a strong separation tendency where highly influential nodes avoid connecting with highly influenced nodes. In contrast, the out-out modality exhibits a sharp increase in positive assortativity as the excess out-strength of nodes rises, suggesting a strong clustering tendency among highly influential nodes connecting with others of similar influence. %These patterns are consistent across all markets, with the FTSE100 displaying the highest positive in-in and out-out assortativity, while the Dow Jones 30 and Eurostoxx 50 exhibit slightly weaker trends.

The patterns across markets consistently reveal meaningful trends in node connectivity under each local assortativity measure. For the Extended Piraveenan et al. (2010) measure, the orange (in-out) and red (out-out) lines rise consistently with increasing node weights, indicating that nodes that are both highly influenced (in) by and highly influence (out) other nodes tend to tie up with nodes that in turn highly influence others (out). Meanwhile, the blue (in-in) and green (out-in) lines consistently decrease, highlighting that nodes that are highly influenced (in) by and highly influence (out) other nodes tend not to tie up with nodes that in turn are highly influenced by others (in).

For the Sabek and Pigorsch (2023) measure, the blue (in-in) and red (out-out) lines consistently rise across markets as node weights increase, meaning that nodes that are highly influenced (in) by others or that highly influence (out) other nodes increasingly tie up with nodes that exhibit the same characteristics. In contrast, the orange (in-out) and green (out-in) lines decrease steadily, reflecting that nodes highly influenced (in) by others or nodes that highly influence (out) others tend not to tie up with nodes that exhibit opposite characteristics (e.g., highly influencing versus highly influenced).

The Peel et al. (2018) measure shows the strongest dynamics, with the red (out-out) line sharply rising across all markets as node weights increase, signifying a strong clustering tendency for nodes that highly influence others to tie up with nodes that in turn highly influence others. Similarly, the blue (in-in) line rises significantly, indicating that nodes highly influenced (in) by others increasingly tie up with similarly influenced nodes. However, the orange (in-out) and green (out-in) lines consistently decrease, reflecting a strong disassortative tendency where nodes that are highly influenced (in) by or highly influence (out) others avoid connecting with nodes that exhibit opposing roles. These consistent patterns across markets highlight the role of increasing node weights in shaping assortative and disassortative behaviors, with clear implications for diversification in portfolio optimization.

These observations provide nuanced insights into the connectivity dynamics within financial networks. As the average excess out-strength of nodes increases, significant variations in assortativity emerge, underscoring the role of influential and highly influenced nodes in shaping network structures and market behaviors. The interplay between node weight, assortativity measures, and modalities emphasizes the importance of understanding local assortativity in financial networks.

\subsection{Empirical findings}\label{results}

Here we present the results of the empirical analysis, applying local assortativity measures (i.e., Extended Piraveenan et al., 2010; Sabek and Pigorsh, 2023; and Peel et al., 2018) to the Dow Jones 30, Euro Stoxx 50, and FTSE 100 indices to assess their impact on portfolio optimization under the Max Quadratic Utility and Max Sharpe methods. The analysis compares Sharpe ratios across different modalities (i.e., in-in, in-out, out-in, and out-out), emphasizing the diversification benefits of network-based metrics relative to the Markowitz benchmark and to the correlation-based local assortativity measures themselves, secondary benchmarks within each measure. The results are broken down into two ways of constructing a portfolio: a weighted and a simple portfolio, equations (\ref{eq:weightedAss}) and (\ref{eq:simpleAss}), respectively.

The out-of-sample analysis for the Dow Jones 30 in Table \ref{tab:dj30_outsample} demonstrates how integrating local assortativity measures into portfolio optimization can enhance returns relative to the benchmark Markowitz approach. The Sharpe ratios across the assortativity modalities and optimization methods provide insights into how network-based metrics are useful for assessing portfolio performance, emphasizing diversification benefits.

\begin{table}
\caption{Out-of-sample results for Dow Jones 30}
\label{tab:dj30_outsample}
\begin{adjustbox}{max width=\textwidth}
%\scriptsize
\begin{tabular}{llp{1.6cm}p{1.6cm}p{1.6cm}lp{1.6cm}p{1.6cm}p{1.6cm}}
\toprule
 &  & \multicolumn{3}{c}{Max Quadratic Utility} && \multicolumn{3}{c}{Max Sharpe} \\\cmidrule{3-5}\cmidrule{7-9}
Local Assortativity Measures &  & Expected Return & Annual Volatility & Sharpe Ratio && Expected Return & Annual Volatility & Sharpe Ratio \\
\midrule
\multicolumn{9}{l}{\textit{Panel A: Weighted Portfolio Assortativity}} \\
 \\
Markowitz Benchmark &  & 0.103 & 0.214 & 0.805 && 0.102 & 0.180 & 0.945 \\
\cline{1-9}
\multirow[c]{5}{*}{Extended Piraveenan et al. (2010)} & Correlation & 0.102 & 0.212 & 0.820 && 0.097 & 0.178 & 0.919 \\
 & in, in & 0.064 & 0.213 & 0.591 && 0.089 & 0.177 & 0.886 \\
 & in, out & 0.121 & 0.215 & 0.872 && 0.109 & 0.178 & 0.945 \\
 & out, in & 0.096 & 0.216 & 0.749 && 0.052 & 0.192 & 0.680 \\
 & out, out & 0.103 & 0.215 & 0.774 && 0.110 & 0.188 & 0.955 \\
\cline{1-9}
\multirow[c]{5}{*}{Sabek and Pigorsch (2023)} & Correlation & 0.104 & 0.214 & 0.806 && 0.105 & 0.182 & 0.955 \\
 & in, in & 0.105 & 0.214 & 0.815 && 0.087 & 0.179 & 0.877 \\
 & in, out & 0.102 & 0.214 & 0.825 && 0.095 & 0.177 & 0.974 \\
 & out, in & 0.105 & 0.215 & 0.790 && 0.087 & 0.186 & 0.887 \\
 & out, out & 0.095 & 0.214 & 0.762 && 0.087 & 0.178 & 0.867 \\
\cline{1-9}
\multirow[c]{5}{*}{Peel et al. (2018)} & Correlation & 0.192 & 0.215 & 1.144 && 0.155 & 0.178 & 1.260 \\
 & in, in & 0.035 & 0.212 & 0.540 && 0.077 & 0.183 & 0.829 \\
 & in, out & 0.079 & 0.216 & 0.786 && 0.080 & 0.174 & 1.052 \\
 & out, in & 0.113 & 0.208 & 0.902 && 0.067 & 0.157 & 0.907 \\
 & out, out & 0.025 & 0.213 & 0.510 && 0.087 & 0.172 & 0.940 \\
\midrule
\multicolumn{9}{l}{\textit{Panel B: Simple Portfolio Assortativity}} \\
 \\
Markowitz Benchmark &  & 0.103 & 0.214 & 0.805 && 0.102 & 0.180 & 0.945 \\
\cline{1-9}
\multirow[c]{5}{*}{Extended Piraveenan et al. (2010)} & Correlation & 0.109 & 0.199 & 0.962 && 0.102 & 0.174 & 1.039 \\
 & in, in & 0.109 & 0.194 & 0.966 && 0.082 & 0.184 & 0.819 \\
 & in, out & 0.121 & 0.181 & 1.187 && 0.118 & 0.174 & 1.098 \\
 & out, in & 0.098 & 0.208 & 0.824 && 0.101 & 0.185 & 0.968 \\
 & out, out & 0.110 & 0.204 & 0.911 && 0.119 & 0.176 & 1.117 \\
\cline{1-9}
\multirow[c]{5}{*}{Sabek and Pigorsch (2023)} & Correlation & 0.117 & 0.208 & 0.920 && 0.109 & 0.175 & 1.048 \\
 & in, in & 0.112 & 0.207 & 0.925 && 0.104 & 0.178 & 0.978 \\
 & in, out & 0.101 & 0.204 & 0.868 && 0.110 & 0.173 & 1.057 \\
 & out, in & 0.102 & 0.205 & 0.870 && 0.115 & 0.173 & 1.105 \\
 & out, out & 0.104 & 0.206 & 0.883 && 0.104 & 0.176 & 1.020 \\
\cline{1-9}
\multirow[c]{5}{*}{Peel et al. (2018)} & Correlation & 0.113 & 0.166 & 1.227 && 0.122 & 0.170 & 1.108 \\
 & in, in & 0.091 & 0.184 & 1.023 && 0.090 & 0.183 & 0.936 \\
 & in, out & 0.099 & 0.168 & 1.151 && 0.116 & 0.170 & 1.165 \\
 & out, in & 0.092 & 0.160 & 1.164 && 0.117 & 0.165 & 1.222 \\
 & out, out & 0.085 & 0.171 & 0.978 && 0.097 & 0.174 & 0.992 \\
\cline{1-9}
\bottomrule
\end{tabular}
\end{adjustbox}
\end{table}

Focusing first on the Weighted Portfolio Assortativity results of Panel A (Table \ref{tab:dj30_outsample}), the Markowitz benchmark Sharpe ratio under the Max Quadratic Utility method is 0.805, while the Max Sharpe optimization achieves a higher ratio of 0.945. For the Extended Piraveenan et al. (2010) measure, the in-out modality stands out, achieving a Sharpe ratio of 0.872 under Max Quadratic Utility and matching the benchmark’s Max Sharpe ratio of 0.945. This suggests that when nodes highly influenced by others tie up with nodes that in turn highly influence others, it contributes to superior portfolio diversification. Similarly, the out-out modality achieves a Sharpe ratio of 0.955 under Max Sharpe optimization, outperforming both the Markowitz benchmark and the same network-based measure computed through the standard correlation matrices. Here, the tendency of highly influential nodes to tie with similarly influential nodes appears to enhance returns while maintaining acceptable risk levels.

Under the Sabek and Pigorsch (2023) measure, the correlation modality (a secondary benchmark within each measure) slightly improves upon the Markowitz benchmark with a Sharpe ratio of 0.806 under Max Quadratic Utility and matches 0.955 under Max Sharpe optimization. The in-out modality produces a Sharpe ratio of 0.825 under Max Quadratic Utility and 0.974 under Max Sharpe optimization, highlighting the diversification benefits when nodes influenced by others connect to influential ones. These results reinforce that disassortativity, where nodes link to those with different characteristics, supports higher risk-adjusted returns, which is a core goal in portfolio optimization. The Peel et al. (2018) measure, instead, shows exceptional performance under Max Sharpe optimization. The correlation modality reaches a Sharpe ratio of 1.260, significantly outperforming the Markowitz benchmark, indicating that this measure captures critical diversification patterns. The out-in modality also achieves strong results, with Sharpe ratios of 0.902 and 0.907 under the two optimization methods, respectively. These results suggest that nodes highly influencing others, yet connecting to highly influenced nodes, play a crucial role in balancing risk and return, further supporting diversification. The in-out modality similarly achieves a higher Sharpe ratio under the Max Sharpe approach (1.052), highlighting the outperformance of network-based assortativity measures against standard portfolio optimization metrics.

Examining the Simple Portfolio Assortativity results of Panel B (Table \ref{tab:dj30_outsample}) reveals further improvements over the Markowitz benchmark. While there are a couple of exceptions, such as the in-in modalities for both the Extended Piraveenan et al. (2010) and the Peel et al. (2018) measures under the Max Sharpe method, all the results demonstrate that the assortativity measures of our models perform better than the benchmark. The Extended Piraveenan et al. (2010) measure achieves the highest Sharpe ratio of 1.187 under the in-out modality with Max Quadratic Utility, exceeding the benchmark by a significant margin. This reflects the value of incorporating connections where highly influenced nodes interact with highly influential ones. Similarly, the out-out modality under Max Sharpe optimization delivers a Sharpe ratio of 1.117, indicating that highly influential nodes connecting with similarly influential nodes enhance portfolio returns. Under the Sabek and Pigorsch (2023) measure, the out-in modality achieves a Sharpe ratio of 1.105 under Max Sharpe optimization, exceeding the benchmark and highlighting the benefits of disassortative connections for portfolio diversification. The Peel et al. (2018) measure continues to deliver strong performance, with the out-in modality reaching a Sharpe ratio of 1.164 under Max Quadratic Utility and an impressive 1.222 under Max Sharpe optimization, underscoring the role of diverse network connections in optimizing risk-adjusted returns.

The expected returns and annual volatility results in Table \ref{tab:dj30_outsample} align closely with the Sharpe ratio findings, providing a consistent narrative for the benefits of network-based portfolio optimization. In Panel A, the Extended Piraveenan et al. (2010) measure under the in-out modality achieves the highest expected return of 0.121 under Max Quadratic Utility, slightly exceeding the Markowitz benchmark of 0.103, while maintaining a comparable annual volatility of 0.215. Similarly, the out-out modality demonstrates competitive expected returns (0.110 under Max Sharpe), with its slightly higher volatility (0.188) offset by superior Sharpe ratio performance. The Sabek and Pigorsch (2023) measure also delivers strong expected returns, particularly under the correlation modality, where it reaches 0.105 under Max Sharpe, marginally outperforming the Markowitz benchmark. The Peel et al. (2018) measure stands out for its exceptional risk-return profile, with the correlation modality achieving a high expected return of 0.192 under Max Quadratic Utility and maintaining a reasonable volatility of 0.215, reinforcing its top-tier Sharpe ratio.

In Panel B, the Simple Portfolio Assortativity results follow similar patterns, with expected returns and volatilities complementing the Sharpe ratio outperformance. For instance, the Extended Piraveenan et al. (2010) measure achieves a notable expected return of 0.121 under Max Sharpe in the out-in modality, coupled with moderate volatility (0.185). Similarly, the Peel et al. (2018) measure’s out-in modality delivers a low volatility of 0.165 under Max Sharpe, supporting its high Sharpe ratio of 1.222. Across both panels, the interplay between expected returns and volatilities substantiates the superior risk-adjusted performance of network-based assortativity measures, emphasizing their practical utility for portfolio optimization.

\begin{table}
\caption{Out-of-sample results for Euro Stoxx 50}
\label{tab:stoxx50_outsample}
\begin{adjustbox}{max width=\textwidth}
%\scriptsize
\begin{tabular}{llp{1.6cm}p{1.6cm}p{1.6cm}lp{1.6cm}p{1.6cm}p{1.6cm}}
\toprule
 &  & \multicolumn{3}{c}{Max Quadratic Utility} && \multicolumn{3}{c}{Max Sharpe} \\\cmidrule{3-5}\cmidrule{7-9}
Local Assortativity Measures &  & Expected Return & Annual Volatility & Sharpe Ratio && Expected Return & Annual Volatility & Sharpe Ratio \\
\midrule
\multicolumn{9}{l}{\textit{Panel A: Weighted Portfolio Assortativity}}  \\
 \\
Markowitz Benchmark &  & 0.061 & 0.229 & 0.670 && 0.067 & 0.195 & 0.794 \\
\cline{1-9}
\multirow[c]{5}{*}{Extended Piraveenan et al. (2010)} & Correlation & 0.062 & 0.229 & 0.678 && 0.068 & 0.196 & 0.800 \\
 & in, in & 0.082 & 0.227 & 0.721 && 0.106 & 0.190 & 0.999 \\
 & in, out & 0.056 & 0.224 & 0.623 && 0.040 & 0.181 & 0.591 \\
 & out, in & 0.065 & 0.230 & 0.704 && 0.121 & 0.196 & 1.019 \\
 & out, out & 0.059 & 0.228 & 0.669 && 0.062 & 0.192 & 0.739 \\
\cline{1-9}
\multirow[c]{5}{*}{Sabek and Pigorsch (2023)} & Correlation & 0.062 & 0.229 & 0.679 && 0.078 & 0.197 & 0.845 \\
 & in, in & 0.056 & 0.229 & 0.638 && 0.065 & 0.194 & 0.772 \\
 & in, out & 0.057 & 0.228 & 0.657 && 0.074 & 0.193 & 0.846 \\
 & out, in & 0.066 & 0.229 & 0.702 && 0.100 & 0.195 & 0.967 \\
 & out, out & 0.059 & 0.228 & 0.661 && 0.078 & 0.192 & 0.821 \\
\cline{1-9}
\multirow[c]{5}{*}{Peel et al. (2018)} & Correlation & 0.089 & 0.232 & 0.689 && 0.093 & 0.188 & 0.797 \\
 & in, in & 0.047 & 0.243 & 0.592 && 0.040 & 0.202 & 0.651 \\
 & in, out & 0.126 & 0.235 & 0.880 && 0.070 & 0.193 & 0.776 \\
 & out, in & 0.043 & 0.224 & 0.543 && 0.013 & 0.175 & 0.524 \\
 & out, out & 0.062 & 0.233 & 0.534 && 0.033 & 0.198 & 0.566 \\
\midrule
\multicolumn{9}{l}{\textit{Panel B: Simple Portfolio Assortativity}}  \\
 \\
Markowitz Benchmark &  & 0.061 & 0.229 & 0.670 && 0.067 & 0.195 & 0.794 \\
\cline{1-9}
\multirow[c]{5}{*}{Extended Piraveenan et al. (2010)} & Correlation & 0.072 & 0.222 & 0.780 && 0.072 & 0.186 & 0.882 \\
 & in, in & 0.071 & 0.198 & 0.844 && 0.073 & 0.192 & 0.825 \\
 & in, out & 0.053 & 0.185 & 0.720 && 0.047 & 0.182 & 0.737 \\
 & out, in & 0.063 & 0.220 & 0.699 && 0.078 & 0.191 & 0.859 \\
 & out, out & 0.066 & 0.220 & 0.722 && 0.066 & 0.187 & 0.825 \\
\cline{1-9}
\multirow[c]{5}{*}{Sabek and Pigorsch (2023)} & Correlation & 0.059 & 0.227 & 0.679 && 0.084 & 0.190 & 0.943 \\
 & in, in & 0.061 & 0.224 & 0.688 && 0.075 & 0.188 & 0.886 \\
 & in, out & 0.066 & 0.223 & 0.718 && 0.073 & 0.187 & 0.860 \\
 & out, in & 0.063 & 0.221 & 0.710 && 0.067 & 0.184 & 0.869 \\
 & out, out & 0.055 & 0.222 & 0.654 && 0.081 & 0.188 & 0.895 \\
\cline{1-9}
\multirow[c]{5}{*}{Peel et al. (2018)} & Correlation & 0.060 & 0.190 & 0.789 && 0.055 & 0.188 & 0.837 \\
 & in, in & 0.067 & 0.196 & 0.778 && 0.048 & 0.192 & 0.706 \\
 & in, out & 0.055 & 0.181 & 0.773 && 0.048 & 0.179 & 0.736 \\
 & out, in & 0.040 & 0.191 & 0.689 && 0.050 & 0.177 & 0.802 \\
 & out, out & 0.079 & 0.185 & 0.899 && 0.069 & 0.182 & 0.872 \\
\cline{1-9}
\bottomrule
\end{tabular}
\end{adjustbox}
\end{table}

The out-of-sample results for the Euro Stoxx 50 in Table \ref{tab:stoxx50_outsample} further demonstrate the usefulness of local assortativity measures on portfolio optimization. Both the Weighted and Simple Portfolio Assortativity panels reveal consistent improvements in Sharpe ratios across various modalities and optimization methods, highlighting the potential of network-based metrics to enhance portfolio performance compared to the Markowitz benchmark.

Focusing on the Weighted Portfolio Assortativity results in Panel A (Table \ref{tab:stoxx50_outsample}), the Markowitz benchmark achieves Sharpe ratios of 0.670 and 0.794 under the Max Quadratic Utility and Max Sharpe optimization methods, respectively. The Extended Piraveenan et al. (2010) measure shows notable improvements, particularly in the out-in modality, where a Sharpe ratio of 1.019 under Max Sharpe exceeds the benchmark. This reflects the benefits of disassortative connections, where stocks that influence others tie with those influenced by others, fostering diversification. The in-in modality also achieves the highest Sharpe ratio of 0.721 under Max Quadratic Utility, outperforming the benchmark and highlighting the value of assortative connections among highly influenced stocks. Expected returns for these modalities remain stable, with the out-in modality under Max Sharpe recording an expected return of 0.121, the highest in this panel. Across the Sabek and Pigorsch (2023) measure, the out-in modality achieves a Sharpe ratio of 0.702 under Max Quadratic Utility and 0.967 under Max Sharpe, reinforcing the diversification benefits of connecting stocks with contrasting characteristics in the financial portfolio network. Peel et al. (2018) delivers the highest Sharpe ratio in this panel, reaching 0.880 under the in-out modality with Max Quadratic Utility, demonstrating the measure's ability to identify diversification opportunities.

In the Simple Portfolio Assortativity results of Panel B (Table \ref{tab:stoxx50_outsample}), the Markowitz benchmark Sharpe ratios remain at 0.670 and 0.794 under the respective optimization methods. However, the assortativity measures consistently surpass these benchmarks. For example, the correlation modality under the Sabek and Pigorsch (2023) measure achieves a Sharpe ratio ranging from 0.860 to 0.895 under Max Sharpe, outperforming the benchmark and underscoring also the value of leveraging standard correlation-based assortativity (with a Sharpe ratio of 0.943). The Peel et al. (2018) measure delivers impressive results, with the out-out modality achieving the highest Sharpe ratio of 0.899 under Max Quadratic Utility, reflecting the benefits of nodes highly influencing others connecting with similarly influential nodes. The Extended Piraveenan et al. (2010) measure also shows strong performance, with the in-in modality achieving a Sharpe ratio of 0.844 under Max Quadratic Utility, highlighting the effectiveness of assortative connections among highly influenced stocks.

The expected returns and annual volatility reported in Table \ref{tab:stoxx50_outsample} closely align with the trends observed in the Sharpe ratios, providing further evidence of the utility of local assortativity measures in portfolio optimization. In Panel A, the Extended Piraveenan et al. (2010) measure demonstrates notable expected returns, with the out-in modality achieving the highest value of 0.121 under Max Sharpe. This result is paired with a stable annual volatility of 0.196, which supports the elevated Sharpe ratio of 1.019. Similarly, the in-in modality under Max Quadratic Utility records an expected return of 0.082 and a slightly lower volatility of 0.227, emphasizing its ability to balance risk and return. For the Sabek and Pigorsch (2023) measure, the out-in modality achieves an expected return of 0.100 under Max Sharpe, while maintaining a volatility of 0.195, reinforcing its role in enhancing risk-adjusted performance. The Peel et al. (2018) measure shows the highest expected return in this panel with the in-out modality under Max Quadratic Utility, reaching 0.126, though this is accompanied by a marginally higher volatility of 0.235.

In Panel B, the Simple Portfolio Assortativity results echo similar patterns. The correlation modality under the Sabek and Pigorsch (2023) measure achieves an expected return of 0.084 with a volatility of 0.190 under Max Sharpe, complementing its strong Sharpe ratio of 0.943. The out-out modality under the Peel et al. (2018) measure delivers the highest expected return of 0.079 under Max Quadratic Utility, coupled with the lowest volatility of 0.185, underscoring its superior risk-return profile. The Extended Piraveenan et al. (2010) measure achieves a balanced outcome, with the in-in modality recording an expected return of 0.071 and a reduced volatility of 0.198 under Max Quadratic Utility, aligning well with its elevated Sharpe ratio of 0.844. Overall, the expected returns and volatility results consistently reinforce the performance of the local assortativity measures, aligning with the Sharpe ratio improvements and demonstrating their robust contribution to portfolio optimization.

\begin{table}
\caption{Out-of-sample results for FTSE 100}
\label{tab:ftse100_outsample}
\begin{adjustbox}{max width=\textwidth}
%\scriptsize
\begin{tabular}{llp{1.6cm}p{1.6cm}p{1.6cm}lp{1.6cm}p{1.6cm}p{1.6cm}}
\toprule
 &  & \multicolumn{3}{c}{Max Quadratic Utility} && \multicolumn{3}{c}{Max Sharpe} \\\cmidrule{3-5}\cmidrule{7-9}
Local Assortativity Measures &  & Expected Return & Annual Volatility & Sharpe Ratio && Expected Return & Annual Volatility & Sharpe Ratio \\
\midrule
\multicolumn{9}{l}{\textit{Panel A: Weighted Portfolio Assortativity}} \\
 \\
Markowitz Benchmark &  & 0.185 & 0.220 & 1.116 && 0.130 & 0.168 & 1.048 \\
\cline{1-9}
\multirow[c]{5}{*}{Extended Piraveenan et al. (2010)} & Correlation & 0.185 & 0.220 & 1.116 && 0.138 & 0.163 & 1.144 \\
 & in, in & 0.175 & 0.217 & 1.068 && 0.108 & 0.166 & 0.896 \\
 & in, out & 0.174 & 0.222 & 0.956 && 0.099 & 0.164 & 0.789 \\
 & out, in & 0.195 & 0.221 & 1.156 && 0.134 & 0.175 & 0.961 \\
 & out, out & 0.185 & 0.220 & 1.095 && 0.123 & 0.169 & 0.918 \\
\cline{1-9}
\multirow[c]{5}{*}{Sabek and Pigorsch (2023)} & Correlation & 0.186 & 0.220 & 1.123 && 0.142 & 0.170 & 1.134 \\
 & in, in & 0.183 & 0.220 & 1.107 && 0.118 & 0.165 & 0.978 \\
 & in, out & 0.185 & 0.220 & 1.123 && 0.121 & 0.166 & 1.002 \\
 & out, in & 0.191 & 0.221 & 1.135 && 0.126 & 0.174 & 0.907 \\
 & out, out & 0.186 & 0.220 & 1.120 && 0.142 & 0.166 & 1.107 \\
\cline{1-9}
\multirow[c]{5}{*}{Peel et al. (2018)} & Correlation & 0.143 & 0.231 & 0.934 && 0.098 & 0.186 & 0.855 \\
 & in, in & 0.147 & 0.227 & 0.786 && 0.098 & 0.180 & 0.692 \\
 & in, out & 0.188 & 0.237 & 1.053 && 0.118 & 0.184 & 0.975 \\
 & out, in & 0.074 & 0.219 & 0.711 && 0.055 & 0.153 & 0.790 \\
 & out, out & 0.105 & 0.224 & 0.790 && 0.112 & 0.168 & 1.021 \\
\midrule
\multicolumn{9}{l}{\textit{Panel B: Simple Portfolio Assortativity}} \\
 \\
Markowitz Benchmark &  & 0.185 & 0.220 & 1.116 && 0.130 & 0.168 & 1.048 \\
\cline{1-9}
\multirow[c]{5}{*}{Extended Piraveenan et al. (2010)} & Correlation & 0.182 & 0.217 & 1.128 && 0.120 & 0.156 & 1.123 \\
 & in, in & 0.110 & 0.164 & 1.153 && 0.102 & 0.160 & 0.959 \\
 & in, out & 0.106 & 0.161 & 1.125 && 0.094 & 0.154 & 0.961 \\
 & out, in & 0.182 & 0.210 & 1.137 && 0.121 & 0.165 & 1.051 \\
 & out, out & 0.178 & 0.213 & 1.111 && 0.112 & 0.159 & 1.041 \\
\cline{1-9}
\multirow[c]{5}{*}{Sabek and Pigorsch (2023)} & Correlation & 0.186 & 0.219 & 1.134 && 0.140 & 0.166 & 1.181 \\
 & in, in & 0.180 & 0.217 & 1.111 && 0.110 & 0.160 & 0.988 \\
 & in, out & 0.183 & 0.215 & 1.123 && 0.118 & 0.160 & 1.070 \\
 & out, in & 0.179 & 0.214 & 1.113 && 0.112 & 0.156 & 1.053 \\
 & out, out & 0.186 & 0.216 & 1.138 && 0.123 & 0.158 & 1.129 \\
\cline{1-9}
\multirow[c]{5}{*}{Peel et al. (2018)} & Correlation & 0.099 & 0.172 & 1.077 && 0.111 & 0.173 & 0.974 \\
 & in, in & 0.062 & 0.162 & 0.810 && 0.086 & 0.162 & 0.837 \\
 & in, out & 0.062 & 0.160 & 0.890 && 0.093 & 0.155 & 0.990 \\
 & out, in & 0.057 & 0.171 & 0.863 && 0.090 & 0.152 & 1.062 \\
 & out, out & 0.073 & 0.164 & 0.946 && 0.102 & 0.155 & 1.046 \\ 
\cline{1-9}
\bottomrule
\end{tabular}
\end{adjustbox}
\end{table}

The out-of-sample results for the FTSE 100, displayed in Table \ref{tab:ftse100_outsample}, provide further evidence of the utility of local assortativity measures in portfolio optimization. The results highlight consistent improvements in Sharpe ratios across different modalities and optimization methods, showcasing the diversification benefits these measures offer compared to the Markowitz benchmark.

Focusing on the Weighted Portfolio Assortativity results of Panel A (Table \ref{tab:ftse100_outsample}), the Markowitz benchmark achieves Sharpe ratios of 1.116 under Max Quadratic Utility and 1.048 under Max Sharpe optimization. The Extended Piraveenan et al. (2010) measure outperforms these benchmarks in key modalities, with the out-in modality delivering the highest Sharpe ratio of 1.156 under Max Quadratic Utility. This indicates that stocks influencing others while being influenced themselves significantly contribute to risk-adjusted returns, enhancing portfolio diversification. The correlation modality under Max Sharpe optimization also performs well, achieving a Sharpe ratio of 1.144, the highest for this optimization method within the measure. Similarly, the Sabek and Pigorsch (2023) measure demonstrates strong performance, with the out-in modality achieving a Sharpe ratio of 1.135 under Max Quadratic Utility, exceeding the benchmark. The correlation and out-out modalities under Max Sharpe optimization achieve the same Sharpe ratio of 1.134, emphasizing the measure’s ability to balance risk and returns through both assortative and disassortative connections. In contrast, the Peel et al. (2018) measure performs more modestly, with the in-out modality delivering its highest Sharpe ratio of 1.053 under Max Quadratic Utility, indicating potential benefits from stocks highly influenced by others connecting with influential stocks.

In the Simple Portfolio Assortativity results of Panel B (Table \ref{tab:ftse100_outsample}), the Markowitz benchmark Sharpe ratios remain at 1.116 and 1.048 under Max Quadratic Utility and Max Sharpe, respectively. The Extended Piraveenan et al. (2010) measure shows competitive performance, with the in-in modality achieving the highest Sharpe ratio of 1.153 under Max Quadratic Utility, suggesting that assortative connections among highly influenced stocks are effective in optimizing returns relative to risk. Similarly, the Sabek and Pigorsch (2023) measure demonstrates strong results, with the out-out modality reaching a Sharpe ratio of 1.138 under Max Quadratic Utility and 1.129 under Max Sharpe optimization, reflecting the benefits of influential stocks connecting with similarly influential ones. Furthermore, the correlation modality under the same measure achieves the highest Sharpe ratio in this panel, reaching 1.181 under Max Sharpe, underscoring the utility of correlation-based assortativity for portfolio performance. The Peel et al. (2018) measure delivers its best result in the out-in modality, achieving a Sharpe ratio of 1.062 under Max Sharpe optimization, demonstrating its ability to leverage disassortative connections for diversification.

The expected returns and annual volatility results in Table \ref{tab:ftse100_outsample} align closely with the Sharpe ratio trends, reinforcing the effectiveness of local assortativity measures in portfolio optimization. In Panel A, the Extended Piraveenan et al. (2010) measure achieves the highest expected return of 0.195 under the out-in modality with Max Quadratic Utility, paired with a moderate volatility of 0.221, supporting its elevated Sharpe ratio of 1.156. Similarly, the correlation modality under Max Sharpe records an expected return of 0.138 and the lowest volatility in this panel at 0.163, contributing to its strong Sharpe ratio of 1.144. The Sabek and Pigorsch (2023) measure delivers robust expected returns, particularly under the out-in modality, which achieves 0.191 with a slightly higher volatility of 0.221 under Max Quadratic Utility. This measure's correlation and out-out modalities under Max Sharpe demonstrate balanced performance, with expected returns of 0.142 and volatilities of 0.170 and 0.166, respectively, highlighting their ability to maintain risk-adjusted returns.

In Panel B, the Simple Portfolio Assortativity results exhibit consistent patterns. The Extended Piraveenan et al. (2010) measure’s in-in modality delivers an expected return of 0.110 with a significantly reduced volatility of 0.164 under Max Quadratic Utility, aligning with its highest Sharpe ratio of 1.153. The Sabek and Pigorsch (2023) measure also performs well, with the correlation modality achieving the highest expected return in this panel at 0.186 under Max Quadratic Utility, coupled with a volatility of 0.219, supporting its exceptional Sharpe ratio of 1.181 under Max Sharpe. The Peel et al. (2018) measure's out-in modality under Max Sharpe optimization delivers a strong performance, with an expected return of 0.090 and the lowest volatility at 0.152, contributing to its Sharpe ratio of 1.062. Overall, these results confirm that the interplay between expected returns and volatilities complements the Sharpe ratio improvements, showcasing the value of local assortativity measures in enhancing portfolio diversification and performance.

A final closer examination reveals that Panel B generally delivers better results than Panel A across all three indices, indicating that summing the individual assortativity of the nodes selected for the portfolio can lead to improved performance. This outcome suggests that constructing portfolios based on the total assortativity of selected nodes, rather than using a weighted average of node assortativities influenced by stock quantities, may be more effective in leveraging the diversification benefits of local assortativity measures. Furthermore, the modalities of in-out and out-in consistently outperform other modalities within each measure across all indices. This consistent pattern underscores the significant role of modalities with contrasting influence factors of nodes, where stocks that are highly influenced connect with highly influential stocks, and vice versa. Such configurations enhance diversification by incorporating assets with complementary characteristics, ultimately resulting in superior risk-adjusted performance. %This insight reinforces the potential of network-based approaches to address portfolio optimization challenges effectively, particularly by emphasizing the critical contribution of contrasting influence dynamics in achieving optimal diversification.

Overall, the results consistently demonstrate that incorporating local assortativity measures enhances portfolio performance compared to the Markowitz benchmark. This improvement is particularly evident in modalities in which nodes connect to those with differing characteristics (i.e., in-out and out-in), supporting greater diversification. Across all measures, both the Max Quadratic Utility and Max Sharpe optimization methods yield higher Sharpe ratios than the Markowitz approach, emphasizing the value of network-based methodologies in portfolio optimization. These findings align with the economic goal of maximizing returns while minimizing volatility, offering a compelling case for integrating local assortativity metrics into financial decision-making.

\section{Conclusion}\label{conclusion}

This study highlights the potential of network-based measures, specifically local assortativity metrics derived from the Mixture Transition Distribution (MTD) model, to enhance portfolio optimization in financial markets. By shifting away from traditional correlation-based networks, which often overlook nonlinear dependencies, our approach leverages directed and weighted financial networks to better represent the intricate interdependencies among financial assets. Through rigorous empirical analysis applied to the constituents of three major indices—the Dow Jones 30, Euro Stoxx 50, and FTSE 100—we demonstrate that portfolios constructed using local assortativity measures consistently outperform those optimized through the classical mean-variance framework of modern portfolio theory. This superior performance, observed across multiple modalities and optimization criteria, underscores the ability of network-based measures to achieve greater risk-adjusted returns while promoting portfolio diversification.

A closer examination of the results reveals that summing the individual assortativity of the nodes selected for the portfolio, as presented in Panel B, generally delivers better performance than the weighted average approach used in Panel A. This finding suggests that constructing portfolios based on the total assortativity of selected nodes, without weighting by stock quantities, more effectively leverages the diversification benefits of local assortativity measures. Additionally, the consistent outperformance of the in-out and out-in modalities within each measure across all indices highlights the significant role of modalities that connect assets with contrasting influence factors. By integrating nodes where highly influenced stocks link to highly influential ones (and vice versa), these configurations foster greater diversification and, in turn, superior risk-adjusted performance.

The study contributes to the literature also by extending the local assortativity framework to directed networks, offering novel insights into the directional influences among assets. Furthermore, our methodology distinguishes itself by adopting optimization standards rooted in modern portfolio theory, such as the Max Quadratic Utility and Max Sharpe ratio methods, rather than value-at-risk constraints. By consistently outperforming these benchmarks, the findings offer actionable insights for practitioners seeking innovative tools for portfolio management, emphasizing the practical utility of network theory in achieving superior financial outcomes.

Despite its contributions, this research is not without limitations. First, while the MTD model effectively captures nonlinear dependencies, its computational complexity may pose challenges for scalability in larger financial networks or datasets at high-frequency levels. Second, the focus on local assortativity metrics may limit the exploration of other network properties that could further enhance portfolio optimization. Future research could address these limitations by developing computationally efficient algorithms for larger networks and incorporating additional network features, such as community structures or centrality measures, into the optimization process. Additionally, extending the analysis to include alternative asset classes or multi-layered financial networks could provide a more comprehensive understanding of network-based portfolio optimization strategies.

Ultimately, this study provides a foundation for integrating advanced network methodologies into financial decision-making, offering a compelling case for their role in optimizing portfolio performance. By bridging theoretical innovation with practical application, the findings pave the way for future exploration into the intersection of network science and finance.

\section*{Acknowledgments}
We acknowledge the comments from conference participants and discussants at the XLVIII annual conference of the Italian Association for Mathematics Applied to Social and Economic Sciences (AMASES) and the Third International Fintech Research Conference.

\section*{Disclosure statement}
We report no potential conflict of interest.

\section*{Disclosure statement}
This study did not receive any grants for funding.

\section*{Data availability statement}
The data that support the findings of this study are not publicly available but can be purchased at \url{https://www.lseg.com/en/data-analytics/products}.

\bibliographystyle{rQUF}
\bibliography{PortfolioMTD}

\clearpage

\appendices
\section{In-sample results}

See Tables \ref{tab:dj30_insample}, \ref{tab:stoxx50_insample}, and \ref{tab:ftse100_insample}. \\

\begin{table}
\caption{In-sample results for Dow Jones 30}
\label{tab:dj30_insample}
\begin{adjustbox}{max width=\textwidth}
%\scriptsize
\begin{tabular}{llp{1.5cm}p{1.6cm}p{1.4cm}p{2cm}lp{1.5cm}p{1.6cm}p{1.4cm}p{2cm}}
\toprule
 &  & \multicolumn{4}{c}{Max Quadratic Utility} && \multicolumn{4}{c}{Max Sharpe} \\\cmidrule{3-6}\cmidrule{8-11}
Local Assortativity Measures &  & Expected Return & Annual Volatility & Sharpe Ratio & Assortativity && Expected Return & Annual Volatility & Sharpe Ratio & Assortativity \\
\midrule
\multicolumn{11}{l}{\textit{Panel A: Weighted Portfolio Assortativity}} \\
 \\
Markowitz Benchmark &  & 0.689 & 0.204 & 3.541 &  && 0.567 & 0.168 & 4.029 &  \\
\cline{1-11}
\multirow[c]{5}{*}{Extended Piraveenan et al. (2010)} & Correlation & 0.683 & 0.202 & 3.544 & -0.001 && 0.438 & 0.172 & 2.737 & 0.003 \\
 & in, in & 0.560 & 0.204 & 2.854 & -0.248 && 0.189 & 0.170 & 1.052 & -0.010 \\
 & in, out & 0.631 & 0.204 & 3.254 & -0.123 && 0.246 & 0.169 & 1.425 & -0.033 \\
 & out, in & 0.681 & 0.205 & 3.481 & -0.027 && 0.379 & 0.181 & 2.191 & -0.010 \\
 & out, out & 0.684 & 0.205 & 3.512 & -0.013 && 0.428 & 0.174 & 2.668 & -0.013 \\
\cline{1-11}
\multirow[c]{5}{*}{Sabek and Pigorsch (2023)} & Correlation & 0.689 & 0.204 & 3.541 & -0.003 && 0.535 & 0.170 & 3.559 & -0.002 \\
 & in, in & 0.686 & 0.204 & 3.538 & -0.000 && 0.492 & 0.167 & 3.344 & 0.002 \\
 & in, out & 0.684 & 0.203 & 3.537 & -0.008 && 0.460 & 0.168 & 3.055 & -0.002 \\
 & out, in & 0.689 & 0.205 & 3.518 & -0.011 && 0.475 & 0.173 & 3.085 & -0.007 \\
 & out, out & 0.686 & 0.203 & 3.543 & -0.005 && 0.473 & 0.167 & 3.192 & -0.005 \\
\cline{1-11}
\multirow[c]{5}{*}{Peel et al. (2018)} & Correlation & 0.489 & 0.218 & 2.303 & -0.636 && 0.179 & 0.178 & 0.875 & 0.022 \\
 & in, in & 0.384 & 0.205 & 2.025 & -0.715 && 0.168 & 0.181 & 0.862 & 0.071 \\
 & in, out & 0.400 & 0.216 & 2.001 & -1.026 && 0.116 & 0.175 & 0.655 & -0.045 \\
 & out, in & 0.339 & 0.206 & 1.694 & -1.319 && 0.058 & 0.156 & 0.333 & -0.338 \\
 & out, out & 0.331 & 0.210 & 1.733 & -1.093 && 0.095 & 0.174 & 0.513 & -0.071 \\
\midrule
\multicolumn{11}{l}{\textit{Panel B: Simple Portfolio Assortativity}} \\
 \\
Markowitz Benchmark &  & 0.689 & 0.204 & 3.541 &  && 0.567 & 0.168 & 4.029 &  \\
\cline{1-11}
\multirow[c]{5}{*}{Extended Piraveenan et al. (2010)} & Correlation & 0.571 & 0.194 & 3.150 & -0.241 && 0.475 & 0.166 & 3.393 & -0.287 \\
 & in, in & 0.473 & 0.187 & 2.747 & -1.381 && 0.474 & 0.176 & 3.110 & -1.390 \\
 & in, out & 0.413 & 0.178 & 2.555 & -1.236 && 0.440 & 0.167 & 3.105 & -1.262 \\
 & out, in & 0.618 & 0.199 & 3.293 & -0.276 && 0.517 & 0.171 & 3.510 & -0.299 \\
 & out, out & 0.613 & 0.196 & 3.313 & -0.211 && 0.494 & 0.167 & 3.477 & -0.249 \\
\cline{1-11}
\multirow[c]{5}{*}{Sabek and Pigorsch (2023)} & Correlation & 0.658 & 0.200 & 3.473 & -0.050 && 0.489 & 0.167 & 3.403 & -0.113 \\
 & in, in & 0.634 & 0.197 & 3.389 & -0.088 && 0.495 & 0.167 & 3.467 & -0.140 \\
 & in, out & 0.621 & 0.197 & 3.361 & -0.146 && 0.480 & 0.165 & 3.443 & -0.191 \\
 & out, in & 0.626 & 0.196 & 3.388 & -0.135 && 0.465 & 0.165 & 3.299 & -0.213 \\
 & out, out & 0.627 & 0.196 & 3.373 & -0.126 && 0.486 & 0.167 & 3.424 & -0.182 \\
\cline{1-11}
\multirow[c]{5}{*}{Peel et al. (2018)} & Correlation & 0.231 & 0.171 & 1.597 & -3.628 && 0.395 & 0.168 & 2.815 & -3.851 \\
 & in, in & 0.373 & 0.178 & 2.345 & -2.725 && 0.446 & 0.175 & 2.955 & -2.675 \\
 & in, out & 0.283 & 0.173 & 1.934 & -4.249 && 0.413 & 0.165 & 2.961 & -4.221 \\
 & out, in & 0.159 & 0.165 & 1.320 & -10.186 && 0.388 & 0.161 & 2.924 & -10.369 \\
 & out, out & 0.297 & 0.171 & 2.024 & -4.953 && 0.425 & 0.168 & 3.009 & -4.891 \\
\cline{1-11}
\bottomrule
\end{tabular}
\end{adjustbox}
\end{table}

\begin{table}
\caption{In-sample results for Euro Stoxx 50}
\label{tab:stoxx50_insample}
\begin{adjustbox}{max width=\textwidth}
%\scriptsize
\begin{tabular}{llp{1.5cm}p{1.6cm}p{1.4cm}p{2cm}lp{1.5cm}p{1.6cm}p{1.4cm}p{2cm}}
\toprule
 &  & \multicolumn{4}{c}{Max Quadratic Utility} && \multicolumn{4}{c}{Max Sharpe} \\\cmidrule{3-6}\cmidrule{8-11}
Local Assortativity Measures &  & Expected Return & Annual Volatility & Sharpe Ratio & Assortativity && Expected Return & Annual Volatility & Sharpe Ratio & Assortativity \\
\midrule
\multicolumn{11}{l}{\textit{Panel A: Weighted Portfolio Assortativity}} \\
 \\
Markowitz Benchmark &  & 0.734 & 0.213 & 3.538 &  && 0.609 & 0.177 & 3.989 &  \\
\cline{1-11}
\multirow[c]{5}{*}{Extended Piraveenan et al. (2010)} & Correlation & 0.733 & 0.213 & 3.537 & 0.002 && 0.564 & 0.179 & 3.503 & 0.003 \\
 & in, in & 0.623 & 0.216 & 2.994 & -0.232 && 0.229 & 0.192 & 1.174 & -0.013 \\
 & in, out & 0.699 & 0.213 & 3.379 & -0.072 && 0.291 & 0.176 & 1.663 & -0.018 \\
 & out, in & 0.727 & 0.214 & 3.495 & -0.020 && 0.434 & 0.190 & 2.446 & -0.007 \\
 & out, out & 0.732 & 0.213 & 3.534 & -0.007 && 0.508 & 0.179 & 3.126 & -0.007 \\
\cline{1-11}
\multirow[c]{5}{*}{Sabek and Pigorsch (2023)} & Correlation & 0.734 & 0.214 & 3.534 & -0.002 && 0.609 & 0.179 & 3.889 & -0.002 \\
 & in, in & 0.733 & 0.213 & 3.534 & -0.001 && 0.570 & 0.178 & 3.637 & -0.000 \\
 & in, out & 0.732 & 0.213 & 3.531 & -0.004 && 0.533 & 0.181 & 3.255 & -0.002 \\
 & out, in & 0.731 & 0.213 & 3.526 & -0.009 && 0.523 & 0.182 & 3.201 & -0.005 \\
 & out, out & 0.732 & 0.213 & 3.539 & -0.003 && 0.562 & 0.177 & 3.581 & -0.004 \\
\cline{1-11}
\multirow[c]{5}{*}{Peel et al. (2018)} & Correlation & 0.590 & 0.222 & 2.678 & -0.452 && 0.211 & 0.184 & 1.056 & -0.009 \\
 & in, in & 0.415 & 0.228 & 1.934 & -0.995 && 0.165 & 0.197 & 0.811 & 0.038 \\
 & in, out & 0.383 & 0.234 & 1.730 & -1.465 && 0.121 & 0.197 & 0.579 & -0.072 \\
 & out, in & 0.403 & 0.219 & 1.918 & -1.246 && 0.067 & 0.172 & 0.351 & -0.330 \\
 & out, out & 0.365 & 0.233 & 1.722 & -1.236 && 0.109 & 0.196 & 0.532 & -0.069 \\
\midrule
\multicolumn{11}{l}{\textit{Panel B: Simple Portfolio Assortativity}} \\
 \\
Markowitz Benchmark &  & 0.734 & 0.213 & 3.538 &  && 0.609 & 0.177 & 3.989 &  \\
\cline{1-11}
\multirow[c]{5}{*}{Extended Piraveenan et al. (2010)} & Correlation & 0.670 & 0.210 & 3.276 & -0.086 && 0.469 & 0.178 & 3.014 & -0.190 \\
 & in, in & 0.433 & 0.195 & 2.313 & -1.656 && 0.462 & 0.185 & 2.842 & -1.678 \\
 & in, out & 0.338 & 0.183 & 1.955 & -1.366 && 0.400 & 0.177 & 2.590 & -1.467 \\
 & out, in & 0.664 & 0.208 & 3.289 & -0.247 && 0.539 & 0.179 & 3.447 & -0.274 \\
 & out, out & 0.671 & 0.208 & 3.328 & -0.141 && 0.508 & 0.175 & 3.314 & -0.188 \\
\cline{1-11}
\multirow[c]{5}{*}{Sabek and Pigorsch (2023)} & Correlation & 0.727 & 0.212 & 3.524 & -0.012 && 0.532 & 0.178 & 3.343 & -0.054 \\
 & in, in & 0.703 & 0.210 & 3.442 & -0.052 && 0.510 & 0.177 & 3.280 & -0.112 \\
 & in, out & 0.686 & 0.210 & 3.368 & -0.106 && 0.502 & 0.177 & 3.229 & -0.159 \\
 & out, in & 0.688 & 0.208 & 3.408 & -0.120 && 0.465 & 0.175 & 3.001 & -0.203 \\
 & out, out & 0.688 & 0.209 & 3.398 & -0.081 && 0.505 & 0.177 & 3.250 & -0.143 \\
\cline{1-11}
\multirow[c]{5}{*}{Peel et al. (2018)} & Correlation & 0.258 & 0.187 & 1.420 & -3.089 && 0.359 & 0.187 & 2.184 & -3.901 \\
 & in, in & 0.321 & 0.195 & 1.759 & -3.823 && 0.440 & 0.188 & 2.670 & -3.839 \\
 & in, out & 0.149 & 0.187 & 1.011 & -5.781 && 0.364 & 0.178 & 2.386 & -6.153 \\
 & out, in & 0.075 & 0.195 & 0.723 & -14.625 && 0.312 & 0.175 & 2.131 & -17.378 \\
 & out, out & 0.198 & 0.189 & 1.298 & -6.763 && 0.393 & 0.178 & 2.565 & -6.833 \\
\cline{1-11}
\bottomrule
\end{tabular}
\end{adjustbox}
\end{table}

\begin{table}
\caption{In-sample results for FTSE 100}
\label{tab:ftse100_insample}
\begin{adjustbox}{max width=\textwidth}
%\scriptsize
\begin{tabular}{llp{1.5cm}p{1.6cm}p{1.4cm}p{2cm}lp{1.5cm}p{1.6cm}p{1.4cm}p{2cm}}
\toprule
 &  & \multicolumn{4}{c}{Max Quadratic Utility} && \multicolumn{4}{c}{Max Sharpe} \\\cmidrule{3-6}\cmidrule{8-11}
Local Assortativity Measures &  & Expected Return & Annual Volatility & Sharpe Ratio & Assortativity && Expected Return & Annual Volatility & Sharpe Ratio & Assortativity \\
\midrule
\multicolumn{11}{l}{\textit{Panel A: Weighted Portfolio Assortativity}} \\
 \\
Markowitz Benchmark &  & 0.883 & 0.211 & 4.221 &  && 0.650 & 0.149 & 5.043 &  \\
\cline{1-11}
\multirow[c]{5}{*}{Extended Piraveenan et al. (2010)} & Correlation & 0.883 & 0.211 & 4.222 & 0.001 && 0.579 & 0.148 & 4.381 & 0.001 \\
 & in, in & 0.826 & 0.210 & 3.962 & -0.105 && 0.212 & 0.158 & 1.293 & -0.001 \\
 & in, out & 0.865 & 0.210 & 4.131 & -0.032 && 0.280 & 0.149 & 1.876 & -0.002 \\
 & out, in & 0.881 & 0.211 & 4.211 & -0.007 && 0.469 & 0.161 & 3.088 & -0.003 \\
 & out, out & 0.883 & 0.211 & 4.214 & -0.001 && 0.538 & 0.151 & 3.922 & -0.001 \\
\cline{1-11}
\multirow[c]{5}{*}{Sabek and Pigorsch (2023)} & Correlation & 0.884 & 0.211 & 4.219 & 0.000 && 0.653 & 0.153 & 4.809 & 0.000 \\
 & in, in & 0.884 & 0.211 & 4.217 & -0.001 && 0.617 & 0.149 & 4.716 & -0.000 \\
 & in, out & 0.883 & 0.211 & 4.221 & -0.001 && 0.591 & 0.151 & 4.374 & -0.000 \\
 & out, in & 0.883 & 0.211 & 4.216 & -0.004 && 0.561 & 0.156 & 4.014 & -0.003 \\
 & out, out & 0.883 & 0.211 & 4.220 & -0.001 && 0.589 & 0.149 & 4.479 & -0.001 \\
\cline{1-11}
\multirow[c]{5}{*}{Peel et al. (2018)} & Correlation & 0.720 & 0.227 & 3.183 & -0.388 && 0.185 & 0.176 & 0.952 & 0.238 \\
 & in, in & 0.413 & 0.229 & 1.875 & -1.232 && 0.100 & 0.177 & 0.535 & 0.018 \\
 & in, out & 0.407 & 0.239 & 1.842 & -1.709 && 0.088 & 0.179 & 0.488 & -0.032 \\
 & out, in & 0.515 & 0.208 & 2.460 & -1.178 && 0.045 & 0.142 & 0.281 & -0.292 \\
 & out, out & 0.400 & 0.228 & 1.802 & -1.380 && 0.072 & 0.167 & 0.383 & -0.034 \\
\midrule
\multicolumn{11}{l}{\textit{Panel B: Simple Portfolio Assortativity}} \\
 \\
Markowitz Benchmark &  & 0.883 & 0.211 & 4.221 &  && 0.650 & 0.149 & 5.043 &  \\
\cline{1-11}
\multirow[c]{5}{*}{Extended Piraveenan et al. (2010)} & Correlation & 0.859 & 0.208 & 4.149 & -0.026 && 0.465 & 0.147 & 3.599 & -0.143 \\
 & in, in & 0.341 & 0.167 & 2.071 & -1.922 && 0.407 & 0.154 & 2.987 & -2.073 \\
 & in, out & 0.344 & 0.162 & 2.153 & -1.370 && 0.342 & 0.150 & 2.568 & -1.783 \\
 & out, in & 0.802 & 0.203 & 3.987 & -0.181 && 0.526 & 0.151 & 3.910 & -0.262 \\
 & out, out & 0.823 & 0.204 & 4.060 & -0.093 && 0.490 & 0.147 & 3.726 & -0.194 \\
\cline{1-11}
\multirow[c]{5}{*}{Sabek and Pigorsch (2023)} & Correlation & 0.881 & 0.211 & 4.208 & -0.003 && 0.562 & 0.153 & 4.038 & -0.041 \\
 & in, in & 0.869 & 0.209 & 4.186 & -0.021 && 0.506 & 0.148 & 3.827 & -0.095 \\
 & in, out & 0.853 & 0.208 & 4.141 & -0.052 && 0.503 & 0.146 & 3.843 & -0.119 \\
 & out, in & 0.850 & 0.207 & 4.131 & -0.076 && 0.460 & 0.146 & 3.462 & -0.173 \\
 & out, out & 0.857 & 0.208 & 4.149 & -0.040 && 0.486 & 0.147 & 3.674 & -0.136 \\
\cline{1-11}
\multirow[c]{5}{*}{Peel et al. (2018)} & Correlation & 0.299 & 0.174 & 1.753 & -3.391 && 0.385 & 0.176 & 2.418 & -4.432 \\
 & in, in & 0.222 & 0.165 & 1.460 & -5.101 && 0.377 & 0.156 & 2.761 & -5.551 \\
 & in, out & 0.171 & 0.165 & 1.261 & -6.542 && 0.301 & 0.153 & 2.302 & -8.084 \\
 & out, in & 0.069 & 0.175 & 0.746 & -16.436 && 0.230 & 0.151 & 1.811 & -25.292 \\
 & out, out & 0.109 & 0.167 & 0.935 & -8.900 && 0.321 & 0.151 & 2.455 & -10.295 \\
\cline{1-11}
\bottomrule
\end{tabular}
\end{adjustbox}
\end{table}

\clearpage

\section{Additional figure}

\begin{figure}
    \centering
    \includegraphics[width=0.5\linewidth]{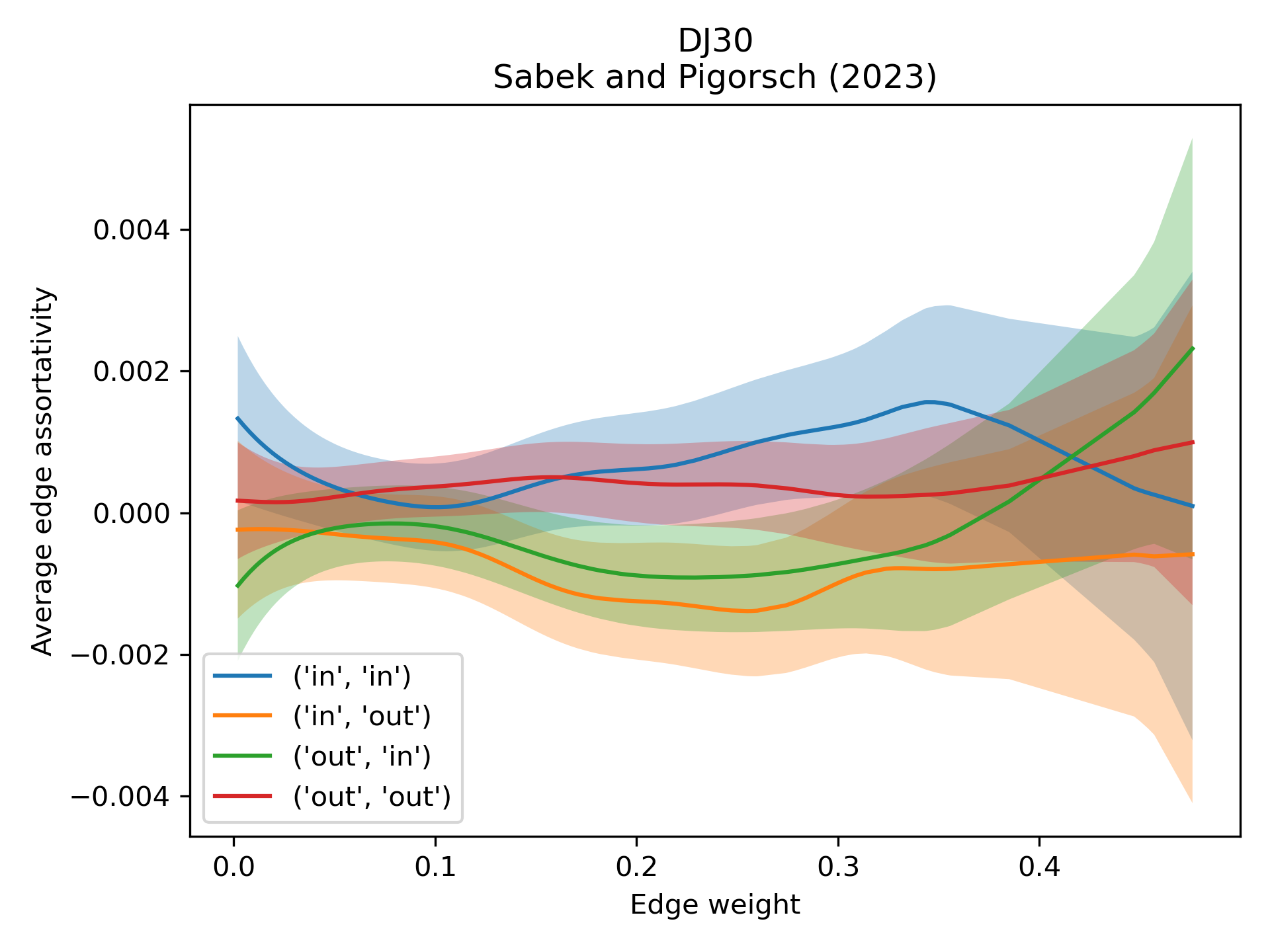}
    \includegraphics[width=0.5\linewidth]{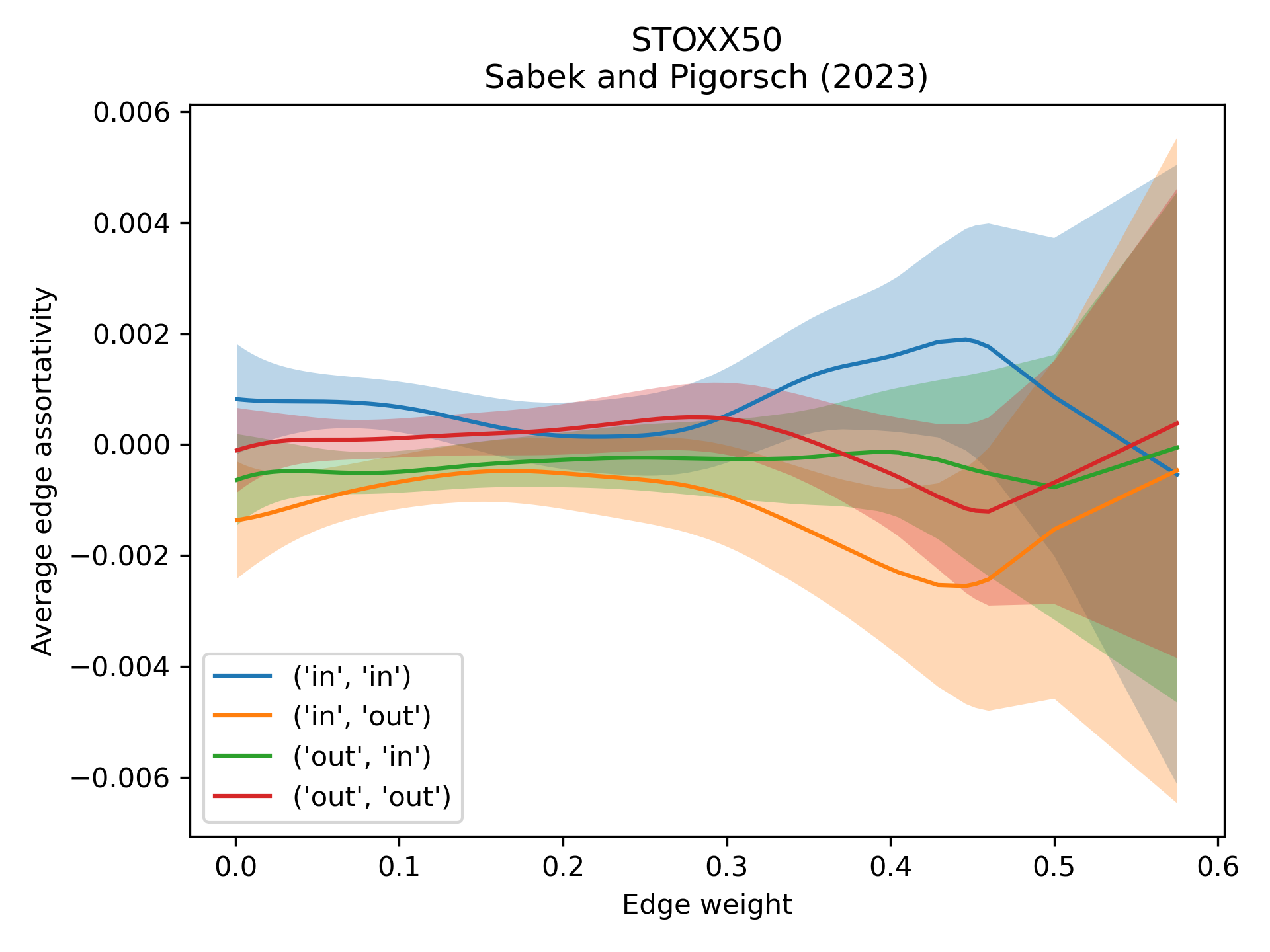}
    \includegraphics[width=0.5\linewidth]{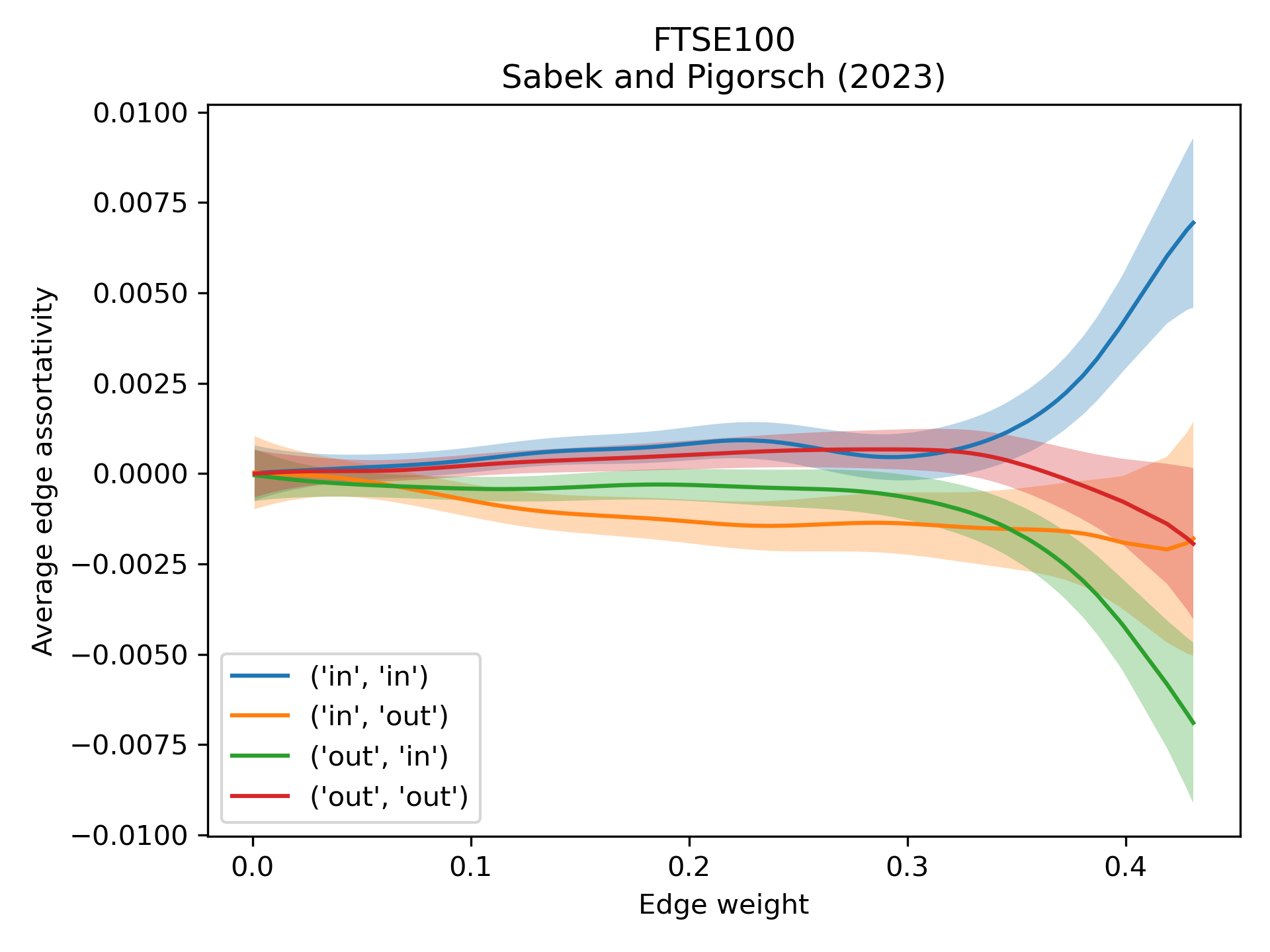}
    \caption{Relationship between edge assortativity and edge weight. The profiles are generated by applying loess regression to smooth the data, with the shaded area representing the 95 per cent confidence intervals.}
    \label{fig2}
\end{figure}

See Figure \ref{fig2}. \\

\end{document}